\documentstyle[11pt,aaspp4]{article}
\received{}
\accepted{}
\input tex.def

\slugcomment{To appear in {\it The Astrophysical Journal}}

\begin{document}

\title{Properties of \hii\ Regions in the Centers of Nearby Galaxies}

\author{Luis C. Ho}
\affil{Department of Astronomy, University of California, Berkeley, CA
94720-3411}

\and

\affil{Harvard-Smithsonian Center for Astrophysics, 60 Garden St., Cambridge,
MA 02138\footnote{Present address.}}

\author{Alexei V. Filippenko}
\affil{Department of Astronomy, University of California, Berkeley, CA
94720-3411}

\and

\author{Wallace L. W. Sargent}
\affil{Palomar Observatory, 105-24 Caltech, Pasadena, CA 91125}

\begin{abstract}

As part of an optical spectroscopic survey of nearby, bright galaxies, we 
have identified a sample of over 200 emission-line nuclei having optical 
spectra resembling those of giant extragalactic \hii\ regions.  Such ``\hii\ 
nuclei,'' powered by young, massive stars, are found in a substantial fraction 
of nearby galaxies, especially those of late Hubble type.  This paper 
summarizes the observational characteristics of \hii\ nuclei, contrasts the 
variation of their properties with Hubble type, and compares the nuclear 
\hii\ regions with those found in galaxy disks.  Similarities and 
differences between \hii\ nuclei and luminous starburst nuclei are 
additionally noted.

Nuclei in early-type spirals (S0--Sbc) on average have low excitation, and 
hence high oxygen abundance (from $\sim$1.1 to 3.3 times the solar 
value), whereas those in late-type systems (Sc--I0) have excitations spanning 
a wide range (corresponding to less than 0.25 to 3.5 times the solar 
oxygen abundance).  The H\al\ luminosities of early-type 
nuclei greatly exceed those of later types.  The enhancement of massive star 
formation may be linked to the higher efficiency with which bars can drive 
gaseous inflow in systems with prominent bulges.  The early-type systems 
also have higher amounts of internal extinction and higher electron densities.

The physical properties of \hii\ nuclei resemble those of giant \hii\ regions
in spiral disks in some ways, but differ in several others.  The two
groups emit comparable H\al\ luminosities and generally have similar 
electron densities.  Because of their unique location in the galaxies, 
nuclear \hii\ regions are characterized by much higher oxygen abundances.  
\hii\ nuclei systematically emit stronger low-ionization forbidden lines than 
disk \hii\ regions, confirming a trend recognized by Kennicutt, Keel, and 
Blaha.  We discuss several possibilities for the origin of the spectral 
variations.

\end{abstract}


\section{Motivation}

The study of extragalactic \hii\ regions provides underpinnings for a number of 
astrophysical problems, such as basic understanding of the formation of 
massive stars, interpretation of the observed abundance patterns in the disks 
of spiral galaxies, and determination of the primordial helium abundance and 
its cosmological implications (Dinerstein 1990; Shields 1990).  The majority of 
observations have concentrated on the most luminous nebulae (so-called 
giant extragalactic \hii\ regions) located either in relatively metal-poor 
dwarf irregular galaxies (e.g., French 1980; Kunth \& Sargent 1983; Terlevich 
\etal 1991) or in the disks of spirals of intermediate to late Hubble type
(e.g., McCall, Rybski, \& Shields 1985).  As the dominant forbidden emission 
lines at optical wavelengths tend to decrease in strength with increasing 
metallicity, comparatively few \hii\ regions have been studied systematically 
in early-type spirals (Oey \& Kennicutt 1993), and fewer still in the central 
regions of galaxies.  Optical emission lines are nearly ubiquitous in 
galactic nuclei (Keel 1983; Ho, Filippenko, \& Sargent 1997b), with the 
majority of the spectra in late-type spirals resembling low-excitation disk 
\hii\ regions (Turnrose 1976; Heckman 1980a; Keel 1983; Ho \etal 1997b).  
Although these ``\hii\ nuclei'' are generally much more feeble in their power 
output than {\it bona fide} starburst nuclei (Balzano 1983), they can serve as 
convenient probes of massive star formation in the 
unique environment of galactic nuclei.  The relatively large number of such 
nuclei in nearby galaxies also permits more detailed inquiries into systematic 
variations of physical properties with host galaxy parameters.  One might 
gain additional insight into the nature of \hii\ nuclei by direct comparison 
with disk \hii\ regions, whose properties are better understood.

The most widely available sources of \hii\ nuclei, until recently, are the 
optical spectroscopic surveys of nearby galaxies published by Heckman, Balick, 
\& Crane (1980), Stauffer (1982), and  Keel (1983).  Kennicutt, Keel, \& Blaha 
(1989; hereafter KKB) compiled all the then available data, supplemented with a 
few new observations, and summarized their salient observed properties.  KKB
found that \hii\ nuclei generally have integrated properties, such as sizes, 
H\al\ luminosities, reddening, and electron densities, comparable to those of 
disk \hii\ regions, although the two classes of nebulae differ in other 
important aspects.

Using a large, homogeneous sample of \hii\ nuclei extracted from a newly 
completed survey, we summarize the statistics of these objects, examine 
systematic variations of observed properties as a function of host galaxy 
type, and draw appropriate comparisons with disk \hii\ regions.  A preliminary 
form of this study was presented in Ho, Filippenko, \& Sargent (1996).

\section{The New Survey}

An optical spectroscopic survey of the nuclei of bright, nearby galaxies 
has recently been completed (Ho, Filippenko, \& Sargent 1995).  High 
signal-to-noise ratio, moderate-resolution (100--200 \kms), long-slit spectra 
were obtained with the Hale 5~m reflector at Palomar Observatory (Filippenko 
\& Sargent 1985).  The sample is defined to be all galaxies listed in the 
Revised Shapley-Ames Catalog of Bright Galaxies (Sandage \& Tammann 1981) with 
$\delta\,>$ 0\deg\ and $B_T\,\leq$ 12.5 mag; it is nearly statistically 
complete and contains 486 galaxies spanning all morphological types.  The 
selection criteria and the relatively large number of objects surveyed ensure 
that the sample provides a fair representation of the local ($z\,\approx\,0$) 
galaxy population, at least for high-surface brightness systems.  The 
observations were taken largely with a 2\asec\ slit, and one-dimensional 
spectra were extracted using an aperture of 2\asec\ $\times$ 4\asec,
corresponding to physical dimensions of $\sim200\,\times\,400$ pc$^2$ for 
the typical distances of the sample galaxies (18 Mpc; Ho, Filippenko, \& 
Sargent 1997a).\footnote{We adopt $H_0$ = 75 \kms\ Mpc$^{-1}$ in this series 
of papers.} The reader may consult Ho \etal (1995) for 
full details of the observations, data reductions, and presentation of 
the spectra, Ho \etal (1997a) for the actual measurements and other 
quantities used in the present analysis, and Ho \etal (1997b) for further 
discussion concerning the statistical properties of the emission-line objects.

Following standard practice (Baldwin, Phillips, \& Terlevich 1981; Veilleux \& 
Osterbrock 1987), we identified the dominant ionization mechanism of each 
nucleus according to the intensity ratios of several prominent optical 
emission lines (Ho \etal 1997a).  \hii\ nuclei are defined to be objects whose 
spectra look similar to those of \hii\ regions, and hence their primary source 
of ionization is assumed to be photoionization by ultraviolet radiation from 
young, massive stars.  Between 4200 and 6900 \AA, the wavelength range 
covered in our survey, the strongest emission lines observed in \hii\ nuclei 
are the hydrogen recombination lines (H$\gamma$, H\bet, and H\al) and
\oiii\ \lamb\lamb 4959, 5007.  The strength of \oiii\ relative to the 
hydrogen lines spans a wide range depending on the excitation of the gas.  The 
low-ionization transitions of \nii\ \lamb\lamb 6548, 6583, 
\sii\ \lamb\lamb 6716, 6731, and especially \oi\ \lamb\lamb 6300, 6363 are 
normally quite weak compared to H\al; indeed, this trait is the primary 
characteristic used to distinguish \hii\ nuclei from active galactic nuclei 
(AGNs).  The sensitivity of our survey, however, is sufficiently high that 
we almost always detect \nii\ and \sii, and we can measure \oi\ in the 
majority ($>$ 80\%) of the objects.  In total, 206 \hii\ nuclei were detected 
in the Palomar survey, of which $\sim$80\% were observed under photometric 
conditions.  

It is important to recognize that our data, as presented here, contain only 
limited spatial information.  Imaging studies of emission-line nuclei (e.g., 
Pogge 1989) have shown that star-forming nuclei often exhibit complicated and 
sometimes extended patterns of line emission.  Without imaging data with 
sufficient angular resolution, it is virtually impossible to define nuclear 
\hii\ regions as single, discreet entities.  The spectra were extracted using 
an aperture of fixed angular size projected onto the sky; this corresponds to 
varying physical dimensions depending on the distance of the galaxy.  
As emphasized by Ho \etal (1997a), the emission-line measurements of 
individual objects in our survey may carry substantial uncertainty.  If the 
line-emitting material is much more extended than our observation aperture, 
we will underestimate the flux.  For more compact emission, the integrated 
light may originate from a number of discrete, and possibly overlapping, 
regions, and the flux for a single region will be overestimated.  On the other 
hand, the {\it statistical} properties of large numbers of objects should be 
much more reliable, as individual fluctuations will average out.  We proceed 
with our analysis in the next section with this principle in mind.

In order to contrast typical \hii\ nuclei with their more luminous 
counterparts, we collected spectroscopic measurements for starburst nuclei from 
the literature.  Objects classified as starburst nuclei generally having 
nuclear H\al\ luminosities greater than 10$^{40}$ \lum\ (e.g., Balzano 1983).  
Thus, in compiling data for the present purpose, we arbitrarily used this 
luminosity cut-off.  We found forbidden-line intensities for 62 objects in
French (1980), Balzano (1983), and Veilleux \& Osterbrock (1987), and H\al\ 
luminosities for 98 objects in Balzano (1983).  Balzano selected her sample 
from the survey of Markarian and his colleagues (Mazzarella \& Balzano 1986, 
and references therein).

Data for disk \hii\ regions are more readily available.  Spectroscopically 
measured line-intensity ratios for about 200 \hii\ regions in spiral disks 
were taken from the studies of McCall \etal (1985) and Ryder (1995), and line 
measurements for a small number of regions in dwarf irregular galaxies were 
used to supplement the low-metallicity end of the \hii-region sequence 
(French 1980; Dinerstein \& Shields 1986).   The study of Kennicutt (1988) 
furnished H\al\ luminosities for 95 first-ranked disk \hii\ regions selected 
from a wide distribution of galaxy morphological types.

\section{Basic Properties of Nuclear \hii\ Regions}

\subsection{Statistics and Host Galaxies}

A detailed discussion of the detection rate of the emission-line nuclei in 
our survey, including that of \hii\ nuclei, can be found in Ho \etal (1997b; 
see also Ho 1996).  In brief, \hii\ nuclei constitute 42\% of all galaxies 
brighter than $B_T$ = 12.5 mag.  Late-type galaxies contain \hii\ 
nuclei much more frequently than early-type galaxies.  Up to 80\% of the 
galaxies with Hubble type between from Sc and Im have an \hii\ classification, 
to be compared with 51\% for Sb, 22\% for Sa, and 8\% for S0.  We found no
ellipticals with an \hii\ nucleus in our sample.

The galaxies hosting \hii\ nuclei tend to be less luminous than those 
hosting AGNs (Ho \etal 1997b), as one would expect from the correlation 
between nuclear type and galaxy morphology.  Our subsequent analysis will 
frequently contrast early-type and late-type systems, where ``early'' here 
refers to types ranging from S0 to Sbc and ``late'' corresponds to Sc through 
I0.  A small number (6) of galaxies with peculiar classifications were omitted.
We show in Figure 1, for later reference, the absolute magnitudes of these 
two groups; as is well known (e.g., Roberts \& Haynes 1994), late-type galaxies
clearly have lower luminosities, although there is considerable overlap 
between the two groups.

Roughly 50\%--60\% of the disk galaxies in the survey are barred, consistent 
with the bar fraction in the general field galaxy population (Ho \etal 1997a).
Considerations of the dynamical influence of a bar on the gaseous component 
of the disk predict that barred systems should show enhanced star formation in 
their centers.  While the incidence of \hii\ nuclei is evidently not higher in 
the expected sense, we do find evidence for an increased star-formation rate 
in the barred sample (Ho, Filippenko, \& Sargent 1997c).  As discussed in 
greater detail in that paper, bars in early-type spirals seem to exert the 
greatest impact on nuclear star formation, while their effect in late-type 
systems is relatively minor.
 
\subsection{H\al\ Luminosity and H\bet\ Equivalent Width}

The H\al\ luminosities\footnote{Recall the caveat mentioned in \S\ 2 
that our emission-line measurements refer only to the limited, 
distance-dependent spatial scale sampled by the narrow slit of the 
observations.  In this comparison, we assume that both the H\al\ luminosities 
and equivalent widths are representative of \hii\ nuclei in a statistical 
sense.  For spatially resolved nebulae, the emission will be underestimated by 
an amount which depends on the spatial variation of the line emissivity.  
Although the individual emission-line parameters are probably inaccurate, it 
is hoped that the ensemble properties will be representative of the class as 
a whole.  The H\al\ luminosities of the starburst nuclei should be fairly 
reliable, since Balzano (1983) selected them based on their semi-stellar 
appearance, and the fluxes were measured through either a 3\asec\ $\times$ 
8\asec\ aperture or a 4\asec\ round aperture.  Similarly, Kennicutt (1988) 
obtained aperture photometry for the first-ranked disk \hii\ regions, which 
should yield accurate integrated H\al\ fluxes.} [L(H\al)] of \hii\ nuclei span 
an enormous range 
(L(H\al) $\approx$ 2\e{36}--4\e{41} \lum; median = 1.6\e{39} \lum), with the 
upper end partly overlapping with starburst nuclei (Fig. 2).  The lower end 
of the luminosity distribution undoubtedly suffers from severe selection 
biases, which themselves must depend on Hubble type, as it becomes 
progressively more difficult to detect weak emission lines superposed on the 
bright stellar background of the bulge.  Compared with late-type (Sc--I0) 
galaxies, early-type (S0--Sbc) systems have nuclei with much higher H\al\ 
luminosities; the increase in the median L(H\al) is about a factor of 9.  The 
cumulative distributions of the two samples differ very significantly 
according to the Kolmogorov-Smirnov (K-S) test (Press \etal 1986); the 
probability ($P_{\rm KS}$) that the two distributions are drawn from the same 
population is 1.9\e{-5}.  Although the median distance of the early-type 
group (21.5 Mpc) is 30\% larger than that of the late-type group (16.9 Mpc),
this effect alone probably cannot account for the large difference in
L(H\al) between the two.  If the surface brightness of H\al\ emission were
constant in the central regions of both groups, L(H\al) in the early-type
galaxies would be only $\sim$70\% higher.  This is a realistic upper limit
to the distance-induced enhancement, since in most cases the H\al\ emission
should be more centrally concentrated.  Instead, the increase of L(H\al)
among early types most likely can be attributed to enhancement of central
star formation resulting from bar-driven gas inflow, an effect observed
{\it only} in early-type systems (Ho \etal 1997c).

First-ranked disk \hii\ regions have luminosities intermediate between
those of early- and late-type \hii\ nuclei, most closely matching nuclei in
Sc--Scd hosts.  But, as found by KKB, \hii\ nuclei on the whole do not differ 
appreciably from disk \hii\ regions in terms of H\al\ luminosity.

We confirm that \hii\ nuclei have much smaller Balmer emission-line equivalent
widths than do disk \hii\ regions (KKB).  [In this comparison we use 
the equivalent width of H\bet\ instead of H\al\ because the latter was not 
given in the literature sources we chose.  EW(H\bet) is not tabulated in our 
data base, but it can be calculated easily from the information given in 
Ho \etal (1997a).]  This primarily underscores the 
difference in the underlying stellar population between the two classes of 
objects, since it has been shown that the emission-line luminosities are 
comparable.  The optical stellar continuum in galactic nuclei predominantly 
comes from red, evolved stars.  With the exception of the nuclei in very 
late-type hosts, the median equivalent width of H\bet\ emission shows 
little variation among Hubble types (Fig. 3), being typically 4--5 \AA.  
The tail in the Sc--I0 distribution, which reaches values as high as 
350 \AA, is due almost entirely to Sd--Im galaxies and closely resembles 
the distribution for disk \hii\ regions.  This is hardly surprising, 
considering that optically distinct ``nuclei'' very rarely exist in galaxies 
of such late Hubble types.  The central regions of such objects are seldom 
well-defined (Ho \etal 1995), and, during the spectral extractions, we usually 
chose the intensity peak closest to the center of the slit to represent the 
``nucleus.''  Undoubtedly many or most such nuclei are counterparts of 
giant \hii\ regions observed in irregular galaxies and in the disks of 
late-type spirals.

The relatively low H\al\ luminosities of \hii\ nuclei suggest that in general 
the centers of nearby galaxies experience only one-hundredth the amount of 
star formation taking place in rarer, more distant starburst nuclei.  
Indeed, the low Balmer-line equivalent widths observed are consistent with 
the picture that these objects have been forming stars essentially at a 
constant rate of a few percent of a solar mass per year over the age of the 
disk (KKB).

\subsection{Reddening}

As emphasized by KKB and readdressed recently by Shields \& Kennicutt (1995),
dust can have a considerable impact on the structure and on the thermal 
and ionization balance of an \hii\ region.  The effect is expected to be 
particularly large in the regime of high (\gax solar) metallicity, which, as 
will be shown below, typifies most galactic nuclei.  Figure 4 illustrates that 
both \hii\ nuclei and disk \hii\ regions suffer
nonnegligible amounts of extinction as gauged by comparison of the observed
and predicted hydrogen Balmer decrement.  With the exception of a small
number of objects having unusually high reddening (IC 10, NGC 891, NGC 1569,
M82), the dispersion in the distribution of reddening is comparable for \hii\
nuclei and disk \hii\ regions.  \hii\ nuclei do have somewhat higher reddening:
the median values of $E(B-V)$ are 0.54, 0.35, and 0.42 mag for early-type,
late-type, and all nuclei combined, respectively.  This is to be compared with
0.29 mag for disk \hii\ regions.  Performing a K-S test on these distributions
indicates that the difference between disk \hii\ regions and late-type nuclei
is insignificant ($P_{\rm KS}$ = 0.29), but highly significant when 
compared to early-type nuclei ($P_{\rm KS}$ = 0.0022).

We draw attention to a minor caveat to the above discussion.  It is possible 
that the apparently higher reddenings in early-type \hii\ nuclei may result 
from an unknown source of systematic error in the measurement of the Balmer 
decrement.  The equivalent widths of the Balmer emission lines are inherently 
small (\S\ 3.2), and accurate measurement is all the more challenging in 
early-type galaxies because of the strong stellar continuum.  The H\bet\ line 
poses more difficulty than H\al\ for a number of reasons (Ho \etal 1997a) and 
therefore it carries a larger uncertainty, although we can think of no obvious
reason why its intensity might have been systematically underestimated in 
early-type galaxies.  As an illustration of the pitfalls of not properly 
accounting for the stellar absorption, we note that the distribution of 
internal reddening in \hii\ nuclei obtained by V\'eron \& V\'eron-Cetty (1986) 
peaks near 0.8 mag, substantially higher than our value of 0.42 mag.  We 
suspect that the discrepancy can be attributed to the fact V\'eron \& 
V\'eron-Cetty did not correct their spectra for starlight contamination.

A systematic overestimate of $E(B-V)$ for early-type nuclei will affect 
the quantitative details of our discussion on H\al\ luminosity (\S\ 3.2), but 
is unlikely to change the general conclusions.  For example, the maximum 
difference between the median $E(B-V)$ of early-type nuclei and disk 
\hii\ regions amounts to 0.25 mag, which translates to a scaling factor 
of 1.8 at H\al.  Similarly, the \hii\ nuclei in early-type galaxies are 
reddened by 0.19 mag more than those in late-type galaxies, translating to a 
maximum enhancement of 60\% in H\al\ luminosity; this is negligible compared 
to the derived factor of $\sim$10 difference in luminosities.

A potentially much more serious uncertainty arises from our lack of knowledge 
of the {\it total} extinction in galactic nuclei.  The conventional 
method of using the Balmer decrement to estimate the optical extinction
assumes a uniform, foreground dust screen.  This is certainly a gross 
oversimplification for real star-forming regions, perhaps all the more so 
for the complex environment of galaxy nuclei.  Thus, the reddenings 
determined from Balmer decrements should be regarded as lower limits to the 
true values.  With the caveat that the presence of dust grains diminishes the 
number of ionizing photons, measurements of the free-free continuum at radio 
wavelengths can in principle provide a more direct measurement of the 
optical depth, and observations indicate that while the optical extinction 
in disk \hii\ regions is usually underestimated using optical techniques, 
the error is generally not too large (Israel \& Kennicutt 1980; Kaufman \etal 
1987; van der Hulst \etal 1988).  Unfortunately, the same technique has not 
been widely applied to galactic nuclei.  

\subsection{Electron Density, Ionized Hydrogen Mass, and Volume Filling Factor}

Nearly all of the \hii\ nuclei in our survey have reliable measurements of 
the density-sensitive \sii\ \lamb\lamb 6716, 6731 doublet.  The lines 
fall on a relatively uncomplicated part of the stellar continuum, and the 
moderately high resolution ($\sim$100 \kms) of our red spectra allows 
accurate deblending of the lines.  The electron densities ($n_e$) were 
derived from the \sii\ \lamb 6716/\lamb 6731 ratio for an electron 
temperature of 10$^4$ K and using the latest atomic parameters for S$^+$ 
(see Ho \etal 1997a).  For consistency we rederived $n_e$ for the disk 
\hii\ regions in the same manner using the published \sii\ intensities.  The 
vast majority of the nuclei have \sii\ ratios in the low-density limit (Fig. 
5).  The average density (180 \cc) and dispersion (200 \cc) for the \hii\ 
nuclei in our sample are quite similar to those of the disk \hii\ regions 
studied by KKB; using the average \sii\ ratio given by McCall \etal (1985), 
for example, we find that disk regions typically have $n_e$ $\approx$ 140 \cc.
KKB, on the other hand, concluded that \hii\ regions in nuclei on average 
have much higher densities than those in disks.  We believe that the 
discrepancy probably can be traced to a subtle systematic error in the 
\sii\ measurements of Stauffer (1982) and Keel (1983), whose surveys KKB 
used to compile much of their sample of \hii\ nuclei.  The source of this 
error has been discussed by Ho (1996).

Interestingly, the electron densities are about a factor of two larger
in early-type hosts than in late-type hosts (median $n_e$ = 180 \cc\
versus 80 \cc).  The K-S test assigns a probability of 0.0016 that the two
distributions are drawn from the same population.  It seems highly unlikely
that this finding is influenced by systematic errors due to measurement.
As stated above, the spectral region concerned generally is unhampered by
strong stellar absorption.  Suppose that \hii\ nuclei in early-type hosts
have lower electron temperatures than those in late-type hosts, as might be
anticipated, for example, from the inverse correlation between metal
abundance and nebular temperature (e.g., Dinerstein 1990).  Indeed, if
we assume a lower value of $T_e$ for early-type nuclei, $n_e$ would decrease, 
but the difference persists at a significant level.  Adopting $T_e$ = 5\e{3} K
for early-type nuclei, for instance, decreases the median density by only 
$\sim$25\%, and the K-S test yields $P_{\rm KS}$ = 0.0079 for the density 
distributions of the two morphological groups.  Thus, it seems 
that \hii\ nuclei in early-type galaxies truly do have higher electron densities
than those in late-type systems.  This might be a natural consequence of the 
different degree of bulge dominance in the two cases.  Gas sitting in the 
deeper gravitational potential of a big bulge naturally is subject to a 
larger pressure which then results in a higher density.  In general, the 
central regions of galaxies are likely to experience much higher 
interstellar pressures than in the disk (Spergel \& Blitz 1992; Helfer \& 
Blitz 1993).  We find it somewhat surprising, therefore, that the densities 
deduced in the nuclear \hii\ regions should be so similar to those in disk 
regions.

Following KKB, we can obtain rough estimates of the ionized hydrogen masses 
by combining the extinction-corrected H\al\ luminosities with the electron 
densities.  $M$(\hii) varies between 10$^2$ to a few\e{6} \solmass, with a 
median value of $\sim$10$^5$ \solmass.  The disk population of giant \hii\ 
regions exhibits similar characteristics (Kennicutt 1984).  The \hii\ masses 
in early-type nuclei are about 2--3 times larger than in late-type nuclei.  
Similarly, the volume filling factors can be crudely 
calculated by assuming that the line-emitting material has dimensions 
comparable to the size of our observing aperture.  This exercise yields 
filling factors ranging from 10$^{-6}$ to 10$^{-1}$, with most of the values 
clustering near 10$^{-5}$; no large variation with Hubble type is seen.  As 
realized by KKB, the ionized gas in nuclear regions evidently attains a much 
higher degree of clumpiness that in the disk.

\subsection{Line-Intensity Ratios}

\subsubsection{[O~III]/H\bet\ Ratio and Oxygen Abundances}

Systematic variations among the emission-line spectra of objects hold 
important clues to their physical conditions.  In this section, we examine 
several well-measured emission-line intensity ratios, paying close attention 
to variations among \hii\ nuclei as a function of Hubble type, between \hii\ 
nuclei and starburst nuclei, and between \hii\ nuclei and disk \hii\ regions.  
In the discussion to follow, we eliminated a small number of highly uncertain 
data points and upper limits.  Our sample is large enough that sufficient data 
remained after this selection process.

Figure 6 illustrates the distributions of \oiii\ \lamb 5007/H\bet\ for the 
different object classes.  \oiii/H\bet\ measures the excitation of the 
nebulae: ``high-excitation'' objects have high \oiii/H\bet, and 
``low-excitation'' objects have low \oiii/H\bet.  Since \oiii\ \lamb 5007 
is an important coolant in \hii\ regions, the ratio is primarily 
sensitive to, and inversely correlated with, the oxygen abundance (e.g., 
McCall \etal 1985; 
Dinerstein 1990).  The \oiii/H\bet\ ratio in \hii\ nuclei spans a wide range 
(Fig. 6{\it a}), forming a spectral sequence mimicking the well-established 
one of giant \hii\ regions (Fig. 6{\it e}).  An obvious difference is that 
\hii\ nuclei cluster toward the low-excitation end of the sequence, since 
galaxy nuclei generally have higher metallicities than disks.  Not 
unexpectedly, the excitation varies with morphological type.  Although 
low-excitation \hii\ regions exist in late-type nuclei (Fig. 6{\it c}), 
high-excitation \hii\ regions are not found in early-type nuclei 
(Fig. 6{\it b}); there appears to be a genuine dearth of high-excitation 
nuclei (\oiii/H\bet\ \gax 2) in S0--Sbc galaxies.  

In the absence of a direct measurement of the electron temperature, it is 
customary to derive the oxygen abundance in an \hii\ region through the use of 
empirically calibrated intensity ratios of bright lines.  A commonly used 
calibrator, introduced by Pagel \etal (1979), is $R_{23}\,\equiv$ (\oii\ + 
\oiii)/H\bet.  Although our spectra do not cover \oii\ \lamb 3727, it is 
well established that in ionization-bounded nebulae \oiii/H\bet\ correlates 
strongly with \oii/H\bet, and hence with $R_{23}$, over a wide range of 
excitation (e.g., McCall \etal 1985; Zaritsky, Elston, \& Hill 1990).  McCall 
\etal show, for instance, that \oiii/H\bet\ closely traces $R_{23}$ for 0.07 
\lax \oiii/H\bet\ \lax 3.7.  Exploiting this empirical fact, we use the 
$R_{23}$--O/H calibration of Edmunds \& Pagel (1984), as updated by McCall 
\etal (1985), to estimate oxygen abundances from our \oiii/H\bet\ 
measurements.  The uncertainty of this calibration, mainly due 
to systematic errors, is estimated to be $\sim$0.2 dex (Edmunds \& Pagel 
1984).  The \hii\ nuclei of early-type spirals have \oiii/H\bet\ ratios 
ranging from $\sim$0.1 to 2.0, with a median value of 0.4.  The corresponding 
range and median value of 12 + log (O/H) are 8.88 to 9.36 and 9.16, 
respectively; for a solar oxygen abundance of 12 + log (O/H) = 8.84 (McCall 
\etal 1985), these translate to 1.1 to 3.3 and 2.1 times the solar value.  
Late-type systems, on the other hand, exhibit \oiii/H\bet\ ranging from as low 
as $\sim$0.06 to as high as 5.5, with a median value of 0.6.  The equivalent 
range in oxygen abundances is probably less than 0.25 to 3.5 times solar, 
and the median value is 1.9 times solar.  The lower bound is rather uncertain
because the $R_{23}$ versus O/H calibration becomes double valued for very 
low values of O/H.


The morphological types of Markarian starburst nuclei are similar to those of
\hii\ nuclei, perhaps with a slightly underrepresentation of galaxies with 
late types (see Fig. 2 of Balzano 1983).  The excitation level of starburst 
nuclei (Fig. 6{\it d}), on the other hand, is strikingly high (contrary to 
Balzano's conclusion) compared to our sample of \hii\ nuclei.  It is tempting 
to interpret this finding in terms of evolution.  Could starburst nuclei
be the predecessors of nearby \hii\ nuclei?  This would be consistent with the 
higher luminosities in starbursts (Fig. 2) and the lower metal abundance 
implied by their higher excitation (median [O/H] $\approx$ [O/H]$_\odot$). 

\subsubsection{``Anomalous'' Strengths of the Low-Ionization Lines}

We next consider the ratio \nii\ \lamb 6583/H\al.  Largely because these two 
lines are often strong, lie close in wavelength, and fall on a sensitive 
portion of modern CCD spectrographs, this ratio is widely used as the 
principal criterion to classify emission-line nuclei.  Because giant \hii\ 
regions virtually never exhibit \nii/H\al\ $>$ 0.6 (e.g., Fig. 8{\it d}), 
it is popular practice (e.g., Keel 1983) to adopt this as the dividing line 
between nuclei photoionized  by stars versus those powered by nonstellar 
processes (i.e., AGNs).  Since we have available other spectral lines 
in addition to these, we prefer to follow the recommendations of Veilleux \& 
Osterbrock (1987) and base our classification also on \sii/H\al\ and 
\oi/H\al\ (Ho \etal 1997a).  We believe this practice to be more reliable, as 
it mitigates possible confusion from selective abundance enhancement of 
nitrogen.  

The distribution of \hii\ nuclei in the \oiii/H\bet\ vs. \nii/H\al\ plane 
parallels the familiar disk \hii\ region sequence, shown in Figure 7 as solid 
points along with the theoretical track of McCall \etal (1985).  For disk 
\hii\ regions, the observed sequence is usually interpreted to reflect 
predominantly the hardening of the ionizing radiation field with decreasing 
metal abundance, although alternative explanations have been advanced 
(cf. McCall \etal 1985; Evans \& Dopita 1985; McGaugh 1991).  
The most striking feature is the clear {\it offset} between the two classes of
objects.  The \hii\ nuclei, especially those with --1 \lax log \oiii/H\bet\
\lax 0.5, have {\it larger} \nii/H\al\ (by about 0.2--0.3 dex)
for the same excitation.  This effect was first noticed by KKB and will be
discussed later.  The range of line ratios is nearly a factor of two
larger among \hii\ nuclei (Fig. 8).  Starburst nuclei have \nii/H\al\ ratios
quite comparable to those of late-type \hii\ nuclei (Fig. 8), and hence they 
are similarly enhanced relative to disk \hii\ regions.  Coziol \etal (1997) 
noticed the same trend in their sample of starbursts.  Figure 9
examines in more detail the dependence of the \nii/H\al\ ratio on Hubble
type.  Early-type systems have the largest \nii/H\al\ ratios, especially
over the range --0.5 \lax log \oiii/H\bet\ \lax 0.2; this may be an indication
that nitrogen is selectively enhanced in these objects. 

As with \nii/H\al, marked differences can be seen between the \sii/H\al\ 
ratios of \hii\ nuclei and disk \hii\ regions (Fig. 10 and 11); the nuclear 
regions show a larger range of observed values, and they are distinctly 
enhanced relative to the disk objects.  The magnitude of the shift is about 
the same as that of \nii/H\al, on the order of 0.2--0.3 dex.  A similar trend 
can be seen qualitatively in \oi/H\al\ (Fig. 13 and 14), although in this case 
there are very few measurements of \oi\ in disk \hii\ regions for comparison 
(the data are virtually non-existent in the low-excitation regime).  
Nevertheless, assuming that the model predictions of McCall \etal (1985) 
adequately represent real \hii\ regions, it appears that the enhancement of 
\oi\ among \hii\ nuclei is even larger than in \nii\ and \sii\ ($\sim$0.5 dex 
for the ridge-line of the \hii\ nuclei population).  One must be cautious, 
however, in drawing conclusions based on observationally unverified models.  
The \oi\ line emits almost exclusively from the partially ionized transition 
region, which, in the case of a nebula photoionized by hot stars, is 
very thin and depends sensitively on the nebular structure.  Campbell (1988) 
finds that her theoretical models of high-excitation \hii\ regions 
systematically underpredict the observed \oi\ strengths by large factors 
(typically on the order of 0.5 dex; see her Fig. 4).   We should also 
remark that \oi/H\al\ has much larger scatter at a given excitation 
than do \nii/H\al\ or \sii/H\al.  Surely measurement errors for this weaker
line must affect the data more severely, but it is conceivable that part of the
larger scatter is real, perhaps reflecting its sensitivity to geometry
and to shock excitation (see \S\ 4.2).

Variations of the \sii/H\al\ ratio with the Hubble type of \hii\ nuclei
(Fig. 12) exhibit a different behavior compared to \nii/H\al.  The nuclei in 
the two morphological groups have essentially the same average \sii/H\al\ 
values at a given excitation, with the late-type objects exhibiting a somewhat 
larger dispersion.  The same trend may also be present in \oi/H\al\ 
(Fig. 13), but the larger scatter and smaller sample size make the comparison 
more difficult.  It is noteworthy that the highest values of \sii\/H\al\ and 
\oi\/H\al\ are associated almost exclusively with late-type galaxies, a point 
we will return to later (\S\ 4.2).  

\section{Discussion}

\subsection{Enhancement of Low-Ionization Lines}

One of the most interesting results of this study is the confirmation of 
KKB's finding that \hii\ nuclei systematically emit stronger low-ionization 
forbidden lines compared to disk \hii\ regions.  Stauffer (1981) first noticed 
this effect for the \nii/H\al\ ratio and suggested shock-wave heating as a 
possible explanation.  The \hii\ nuclei in the study of Veilleux \& Osterbrock 
(1987), on the other hand, did not show any obvious differences with 
disk \hii\ regions, probably because of the small size of the sample.

KKB argued that the most likely explanation for those nuclei showing 
enhancement in the low-ionization species is that they contain a weak AGN.
AGNs have harder ionizing spectra than O and B stars; for a cloud optically 
thick to the Lyman continuum, the high-energy photons create an extensive 
partially ionized zone from which low-ionization transitions such \oi, \sii, 
and, to a lesser extent, \nii\ originate.  As supporting evidence for a 
secondary ionization component from an AGN, KKB showed examples whose 
kinematic properties indicate a composite nature, although it remains to be 
established how common these sources are among objects classified as \hii\ 
nuclei.  Many genuine composite nuclei are known (e.g., V\'eron \etal 
1981; Heckman \etal 1983; Keel 1984; Shields \& Filippenko 1990), and, in 
fact, objects with optical spectra intermediate between those of AGNs and 
\hii\ nuclei constitute a non-trivial fraction of all emission-line nuclei 
(Ho, Filippenko, \& Sargent 1993; Ho 1996; Ho \etal 1997b).  These 
``transition objects'' generally have stronger low-ionization lines 
than do the objects considered here, but it is possible that an AGN component, 
whose strength spans a wide dynamic range, is present in {\it most} galaxy 
nuclei.  Indeed, the formal dividing line between \hii-like and AGN-like 
sources strictly speaking has no physical significance.  The classification 
system for narrow emission-line objects (e.g., Baldwin \etal 1981; Veilleux 
\& Osterbrock 1987; Ho \etal 1997a) is based on a set of empirical, largely 
{\it ad hoc} criteria.  In this interpretation, those \hii\ nuclei showing 
stronger than usual (i.e., compared to disk \hii\ regions) low-ionization 
lines are simply the ones with a weak AGN component.  It would be difficult 
to distinguish this hypothesis from other possibilities (see below) based on 
optical data alone.  If present, the weak active nucleus may be revealed by 
sensitive observations at radio, ultraviolet, or X-ray wavelengths.

In connection with the hidden-AGN hypothesis, we recall that the nuclei 
with the most pronounced \nii\ enhancement appear in early-type galaxies, 
precisely the variety that preferentially host definitive AGNs (Ho \etal 
1997b).  While this may be taken as additional evidence that a nonstellar 
source underlies the observed excitation patterns, we hasten to add 
that the \sii\ and \oi\ lines show just the opposite effect (they are stronger
in late-type galaxies).  Nitrogen might be selectively enriched in the centers
of bulge-dominated galaxies.

Strong low-ionization lines also can be produced through collisional excitation 
by shocks.  KKB considered shocks from supernova remnants (SNRs) an improbable
explanation for the line-strength anomalies, as the number of SNRs
required was deemed excessive, and, in any case, implies that 
\hii\ regions in galactic nuclei differ in fundamental ways from those in 
disks.  However, it does not seem unreasonable that shocks resulting from the 
{\it bulk} or {\it turbulent} motions of the line-emitting gas will give 
rise to enhanced low-ionization line strengths.  The gas velocities in galactic 
nuclei are almost certainly supersonic and larger than those within 
individual, isolated disk \hii\ regions, and noncircular motions, which 
can lead to cloud-cloud collisions, are not uncommon (e.g., Kenney 1996, 
and references therein).  A cursory examination of the line intensities 
predicted for low-velocity shocks (e.g., Shull \& McKee 1979; Ho \etal 1993) 
indicates that the observed ``excess'' can be easily accommodated.  

The photoionization calculations of Filippenko \& Terlevich (1992) and 
Shields (1992) demonstrated that O stars with high effective temperatures 
($T_{eff}$ = 45,000--50,000 K) in an environment with lower than typical  
ionization parameter can generate strong low-ionization lines.  In fact, 
these investigators sought to explain low-ionization nuclear emission-line 
regions (LINERs; Heckman 1980b) with such models.  The presence of high 
effective stellar temperatures in the metal-rich environments of galactic 
nuclei ostensibly violates the known inverse correlation between $T_{eff}$ 
and metallicity obeyed by disk \hii\ regions (Shields \& Tinsley 1976; McCall 
\etal 1985; V\'\i lchez \& Pagel 1988), but Shields (1992) and Filippenko \& 
Terlevich (1992) speculated that nuclear \hii\ regions may depart from 
this trend.  That \hii\ nuclei are required in this scenario to have 
systematically lower ionization parameters than disk \hii\ regions may be
consistent with our finding that the nuclear regions tend to have lower volume 
filling factors (\S\ 3.4).  The ionization parameter, $U$, is proportional 
to $[Q({\rm H}) n f^2]^{1/3}$, where $Q({\rm H})$ is the number of ionizing 
photons s$^{-1}$, $n$ is the gas number density, and $f$ is the volume filling 
factor.  All else being equal, a smaller $f$ leads to a smaller $U$.

Examination of the impact of dust grains on the thermal properties of \hii\ 
regions offers perhaps a more straightforward explanation.  Although Mathis 
(1986) concluded that dust has a minimal effect on the emergent optical 
spectrum of \hii\ regions, the calculations of Shields \& Kennicutt (1995) 
indicate that the influence can be quite appreciable under conditions 
characterized by higher than solar metallicity, as is the case for many \hii\ 
nuclei 
(\S\ 3.5.1).  Compared to the dust-free case, grains enhance optical forbidden 
emission by increasing the equilibrium electron temperature, principally 
through depletion of gas-phase coolants.  We anticipate, therefore, that at 
least part of the spectral differences observed between disk and nuclear 
\hii\ regions might be due to the effect of dust grains, so long as the 
higher metallicity in the nuclear \hii\ regions is also accompanied by a 
higher dust content.  After comparing the results of the Shields \& Kennicutt 
photoionization models with our data (Fig. 7, 10, and 13), it is apparent that 
the predicted line strengths do indeed provide a reasonably good match to the 
observations.  Shields \& Kennicutt performed the same exercise using the 
smaller data set of KKB (which did not contain measurements of the \oi\ line) 
and arrived at the same conclusion.  

Shields \& Kennicutt speculated that the higher dust content inferred in 
galaxy nuclei might be achieved if the gas density in these environments is 
elevated.  The growth of grains appears to be favored in denser media, since 
the level of depletion of many metals increases with increasing gas density 
(e.g., Phillips, Gondhalekar, \& Pettini 1982).  Although higher gas densities 
in the central regions of galaxies may result directly from the high 
interstellar pressures present, there is no evidence of this in our data 
(\S\ 3.4).  Of course, one might question whether the density of the electrons 
derived from the \sii\ lines actually trace the density of the neutral 
material.  The molecular clouds could have substantially higher densities than 
the ionized gas associated with the \hii\ regions.  Alternatively, the dust 
content in galactic nuclei might be enhanced as a natural consequence of the 
higher stellar density present, since dust forms predominantly in evolved 
stars (Mathis 1990).  This scenario can account for the difference in 
internal extinction between nuclear and disk \hii\ regions, as well as that
between early-type and late-type \hii\ nuclei (\S\ 3.3).

Finally, we mention the likelihood of a significant contribution to the line 
emission from the
``diffuse ionized gas.'' This component of interstellar gas, also called the 
``warm ionized medium,'' has long been recognized in the Galaxy (see Reynolds 
1990) and is now known to be pervasive in the disks of other spirals and 
irregulars (e.g., Rand 1996; Greenawalt, Walterbos, \& Braun 1997, and 
references therein).  The optical spectrum of the diffuse gas exhibits
characteristically low stages of ionization as indicated by strong \nii\ and 
\sii, but \oiii\ and \oi\ are generally very weak (e.g., Domg\"{o}rgen \& 
Mathis 1994).  Thus, while the high \nii/H\al\ and \sii/H\al\ ratios in 
the nuclei mimic some of the spectral signatures of the diffuse ionized gas, 
the accompanying high values of \oi/H\al\ are inconsistent with such an 
origin.  The weak \oi\ line in the diffuse emission is thought to reflect 
the matter-bounded conditions of the emitting material (Domg\"{o}rgen \&
Mathis 1994).  The tight correlation of the \sii\ and \oi\ line strengths 
in our nuclei (Fig. 15; see also Fig. 14 in Ho \etal 1997a for a similar 
correlation between \nii\ and \oi), and, indeed, the mere existence of a 
well-defined spectral sequence in the line-intensity ratio diagrams shown 
in Figures 7, 10, and 13 implies that ionization-bounded conditions must 
prevail in nuclear \hii\ regions, as is the case in most disk \hii\ 
regions (McCall \etal 1985).  We therefore think that it is highly unlikely 
that the observed line-ratio anomalies stem from significant contamination 
by the diffuse ionized gas.  Lehnert \& Heckman (1994) noticed that
the integrated spectra of spiral galaxies in the spectrophotometric atlas 
of Kennicutt (1992) {\it also} show enhanced \nii/H\al\ and \sii/H\al\ with 
respect to the standard \hii-region sequence, and they postulated that a 
sizable fraction of the integrated emission could arise from the diffuse 
medium.  They did not, however, present data for \oi, most likely because of 
the difficulty in measuring this weak line in the available data.  Future 
quantification of the \oi\ strength in the integrated optical spectra of 
galaxies will be useful for constraining the nature of their line emission.

\subsection{Evidence for Shock Excitation}

Some \hii\ nuclei in late-type hosts appear to have stronger \sii\ and \oi\ 
lines than nuclei of similar excitation in spirals of earlier type 
(\S\ 3.5.2).  Unusually strong \sii\ emission has previously been noticed by 
Peimbert \& Torres-Peimbert (1992) in some ``\hii\ galaxies.''  Inspecting the 
subsample of 15 nuclei with \sii/H\al\ $\geq$ 0.5 (Fig. 11), it is apparent 
that the majority of them have the highest \oi/H\al\ ratios in the sample 
(median value = 0.054).  This is to be 
expected, since the two line ratios correlate well with each other (Fig. 15), 
reflecting their common emission region.  Most of the objects have Hubble 
types ranging from Scd to Sdm, have rather low luminosities (median 
$M^0_{B_T}$ = --19.4 mag), and are located closer than average (median 
distance = 11.5 Mpc).  Shocks from embedded SNRs, or perhaps 
other forms of mechanical energy injection such as stellar winds from massive 
stars, are probably responsible for the enhanced \sii\ and \oi\ emission in 
these systems, although it is conceivable that photoionization from very 
hot stars can achieve the same effect (see \S\ 4.1).  In view of the low 
metallicities anticipated in such late-type nuclei, the effects of 
dust alone are probably insufficient.  The presence of strong \oi\ can further 
rule out the diffuse ionized gas as the main culprit.  The shock hypothesis 
can be tested either by directly measuring a temperature-sensitive line such 
as \oiii\ \lamb 4363, or by attempting to detect the SNRs themselves.

Perhaps the clearest example of a shock-excited object in our sample 
is IC 2574, which appears as an outlier in Figures 7 and 9.  The very low 
excitation of the spectrum (\oiii/H\bet\ = 0.23) is very peculiar for a 
galaxy classified as a Magellanic spiral (SABm; Ho \etal 1997a).  Indeed, 
Miller \& Hodge (1996) have shown that the oxygen abundance of several of 
its bright \hii\ regions is only about one-tenth solar, 
consistent with the late Hubble type of the galaxy.  The three \hii\ regions 
analyzed by Miller \& Hodge, which do not coincide with the positions sampled 
by our slit, have \oiii/H\bet\ ratios between 3 and 5.  The strengths of 
\nii\ (\nii/H\al\ = 0.07) and \sii\ (\sii/H\al\ = 0.25) in our spectrum also 
appear enhanced by about a factor of 2 to 3 compared to the regions in Miller 
\& Hodge.  Inspection of four other emission regions intersected by our 
slit, which was 2\amin\ long, roughly centered on and oriented along the major 
axis of the galaxy, reveals that one other location (hereafter called ``B'') 
has a spectrum similar to that of the primary peak  (hereafter called ``A,'' 
whose spectrum is plotted in Fig. 29 of Ho \etal 1995).  B is 
separated from A by 31\asec.  The remaining three peaks have spectra similar 
to those of the \hii\ regions studied by Miller \& Hodge, the main difference 
being that \sii\ is somewhat enhanced in our cases.  The simplest explanation 
for the unusual character of spectra A and B is that the emission sampled by 
our slit at these locations comes {\it not} predominantly from photoionized 
gas, but rather from shocked plasma.   The other three \hii\ region-like 
objects, then, owe their enhanced \sii\ emission to a mixture of photoionized 
and shocked gas.  In regions A and B, the simultaneous presence of weak \oiii\ 
and moderately strong \nii\ and \sii\ conforms with the predicted signature of 
collisional ionization in a metal-poor medium (e.g., Dopita \etal 1984).  
\oi\ \lamb 6300 should also be strong, but unfortunately this line is 
corrupted by night-sky emission (the galaxy has a small radial velocity).  
Spectra A and B in IC 2574, in fact, look remarkably like those of the 
SNRs in two other metal-poor irregular galaxies, IC 1613 and 
NGC 6822 (Dopita \etal 1984).  Plotting our measurements of 
\oiii\ \lamb\lamb 4959, 5007/H\bet\ and \nii\ \lamb\lamb 6548, 6583/H\al\ on 
Figure 10 of Dopita et al., we deduce that regions A and B have abundances of 
O/H $\approx$ 0.7\e{-4} and 1.5\e{-4}, respectively, and O/N $\approx$ 12 and 
24, respectively.  Using \nii\ \lamb\lamb 6548, 6583/H\al\ and 
\sii\ \lamb 6731/H\al\ and Figure 8 of Dopita et al. gives somewhat 
different values of O/H $\approx$ 0.2\e{-4} and O/N $\approx$ 5 for both 
regions.  However, if the abundance ratio of sulfur to oxygen were to be 
increased by a factor of 2, which is within the range of variation observed 
in SNRs (Dopita \etal 1984), O/H and O/N would be higher, and the abundances 
deduced from the two diagnostic diagrams would be in agreement with each 
other.  The oxygen and nitrogen 
abundances derived from the shocked regions in IC 2574 agree very well with 
Miller \& Hodge's determination based on the analysis of its \hii\ regions 
(O/H $\approx$ 1.4\e{-4} and O/N $\approx$ 30), and are consistent with 
abundances established in late-type galaxies (e.g., Garnett 1990).

\section{Summary}

\hii\ regions are frequently found in the central few hundred parsecs of 
disk galaxies.  A substantial fraction (42\%) of all spirals in a 
magnitude-limited survey of nearby galaxies contain ``\hii\ nuclei,'' with 
the detection rate of these objects rising steeply toward galaxies with late 
Hubble types. 

The physical properties of \hii\ nuclei in some respects resemble those of 
giant \hii\ regions in spiral disk, but differ in several others.  The 
two classes of \hii\ regions generally have similar H\al\ luminosities, 
electron densities, and ionized hydrogen masses, but \hii\ nuclei are 
characterized by higher oxygen abundances, somewhat higher internal 
extinctions, and much smaller volume filling factors.  The emission-line 
equivalent widths of typical \hii\ nuclei are also very small because of the 
strong dilution from the composite, and generally much older, stellar 
population of the nuclei.

As first recognized by Kennicutt \etal (1989), \hii\ nuclei distinguish 
themselves spectroscopically from disk \hii\ regions by their stronger 
low-ionization forbidden emission.  We verified this trend for the 
\nii\ \lamb\lamb 6548, 6583 and \sii\ \lamb\lamb 6716, 6731 lines, and, for 
the first time, also for the weak \oi\ \lamb\lamb 6300, 6364 transitions.  
There are several possible explanations for the enhancement of the 
low-ionization lines in nuclear \hii\ regions.  These include: (1) a secondary 
ionization source from a weak AGN, (2) exceptionally hot stars coupled with 
nebular conditions in the \hii\ regions that are unique to the nuclear 
environment, (3) shocks, and (4) the modification of the thermal properties 
of metal-enriched gas by dust grains.  A dominant contribution to the 
excess low-ionization emission from the diffuse ionized medium seems 
implausible given the relatively large strength of \oi.

A number of properties of \hii\ nuclei systematically depend on the 
morphological type of the host galaxy, with the dividing line between 
``early'' and ``late'' types occurring near Sbc.  The most obvious 
difference between the two groups is their oxygen abundance, the primary 
parameter controlling the excitation of the line-emitting gas: early-type 
nuclei on average have low excitation, whereas late-type nuclei have 
excitations spanning the full observed range.  The oxygen abundance inferred 
from the \oiii/H\bet\ ratios indicate [O/H] $\approx$ (1.1--3.3) 
[O/H]$_{\odot}$ for early-type systems and [O/H] ranging from \lax 0.25 to 3.5 
[O/H]$_{\odot}$ for late-type systems.  It is interesting to note 
that Markarian starburst nuclei generally have rather high excitation, 
perhaps an indication that these nuclei are experiencing one of their 
earliest generations of vigorous star formation.  A small number of \hii\ 
nuclei, all in early-type hosts, emit exceptionally strong \nii\ lines; 
nitrogen may be selectively enhanced in these objects.  Early-type nuclei have 
much higher H\al\ luminosities, and hence formation rates of massive stars, 
than do late-type nuclei.  The underlying physical basis for this difference 
may be linked to the increased effectiveness of bars in driving gaseous inflow 
in systems with large bulge-to-disk ratios.  The early-type systems are 
additionally more highly reddened, and they possess higher electron 
densities.  Finally, a minority of the late-type spirals contain \hii\ nuclei 
with exceptionally strong \sii\ and \oi\ emission.  We suggest that 
shock-wave heating contributes to the excitation in these regions.

\acknowledgments

The research of L.~C.~H. is currently funded by a postdoctoral fellowship
from the Harvard-Smithsonian Center for Astrophysics.  Financial support for
this work was provided by NSF grants AST-8957063 and AST-9221365, as well as
by NASA grants AR-5291-93A and AR-5792-94A from the Space Telescope Science
Institute (operated by AURA, Inc., under NASA contract NAS5-26555).
L.~C.~H. is grateful to Joe Shields and Ian Evans for informative 
correspondence on \hii\ region models and for providing electronic tabulations 
of their photoionization calculations.  We thank Chris McKee and Hy Spinrad 
for their critical reading of an earlier draft of the manuscript, and an 
anonymous referee for several insightful recommendations.

\clearpage

\centerline{\bf{References}}
\medskip

\refindent
Baldwin, J.~A., Phillips, M.~M., \& Terlevich, R. 1981, \pasp, 93, 5

\refindent
Balzano, V.~A. 1983, \apj, 268, 602

\refindent
Campbell, A. 1988, \apj, 335, 644

\refindent
Coziol, R., Demers, S., Barn\'eoud, R., \& Pe\~{n}a, M. 1997, \aj, in press

\refindent
Dinerstein, H.~L. 1990, in The Interstellar Medium in Galaxies, ed. H.~A.
Thronson, \& J.~M Shull (Dordrecht: Kluwer), 257

\refindent
Dinerstein, H.~L., \& Shields, G.~A. 1986, \apj, 311, 45

\refindent
Domg\"{o}rgen, H., \& Mathis, J.~S. 1994, \apj, 428, 647

\refindent
Dopita, M.~A., Binette, L., D'Odorico, S., \& Benvenuti, P. 1984, \apj, 276,
653

\refindent
Edmunds, M.~G., \& Pagel, B.~E.~J. 1984, \mnras, 211, 507

\refindent
Evans, I.~N., \& Dopita, M.~A. 1985, \apjs, 58, 125

\refindent
Filippenko, A.~V., \& Sargent, W.~L.~W. 1985, \apjs, 57, 503

\refindent
Filippenko, A.~V., \& Terlevich, R. 1992, \apj, 397, L79

\refindent
French, H.~B. 1980, \apj, 240, 41

\refindent
Garnett, D.~R. 1990, \apj, 363, 142

\refindent
Greenawalt, B., Walterbos, R.~A.~M., \& Braun, R. 1997, \apj, in press

\refindent
Heckman, T.~M. 1980a, \aa, 87, 142

\refindent
Heckman, T.~M. 1980b, \aa, 87, 152

\refindent
Heckman, T.~M., Balick, B., \& Crane, P.~C. 1980, \aas, 40, 295

\refindent
Heckman, T.~M., van Breugel, W.~J.~M., Miley, G.~K., \& Butcher, H.~R. 1983,
\aj, 88, 1077

\refindent
Helfer, T.~T., \& Blitz, L. 1993, \apj, 419, 86

\refindent
Ho, L.~C. 1996, in The Physics of LINERs in View of Recent Observations, ed.
M. Eracleous, et al. (San Francisco: ASP), 103

\refindent
Ho, L.~C., Filippenko, A.~V., \& Sargent, W.~L.~W. 1993, \apj, 417, 63

\refindent
Ho, L.~C., Filippenko, A.~V., \& Sargent, W.~L.~W. 1995, \apjs, 98, 477

\refindent
Ho, L.~C., Filippenko, A.~V., \& Sargent, W.~L.~W. 1996, in The
Interplay Between Massive Star Formation, the ISM and Galaxy Evolution, ed.
D. Kunth, et al. (Gif-sur-Yvette: Editions Fronti\`eres), 341

\refindent
Ho, L.~C., Filippenko, A.~V., \& Sargent, W.~L.~W. 1997a, \apjs, in press

\refindent
Ho, L.~C., Filippenko, A.~V., \& Sargent, W.~L.~W. 1997b, \apj, in press

\refindent
Ho, L.~C., Filippenko, A.~V., \& Sargent, W.~L.~W. 1997c, \apj, in press

\refindent
Israel, F.~P., \& Kennicutt, R.~C. 1980, Astron. Lett., 21, 1

\refindent
Kaufman, M., Bash, F.~N., Kennicutt, R.~C., \& Hodge, P.~W. 1987, \apj, 319, 61

\refindent
Keel, W.~C. 1983, \apjs, 52, 229

\refindent
Keel, W.~C. 1984, \apj, 282, 75

\refindent
Kenney, J.~D. 1996, in IAU Colloq. 157, Barred Galaxies, ed. R. Buta, B.~G.
Elmegreen, \& D.~A. Crocker (San Francisco: ASP), 150

\refindent
Kennicutt, R.~C. 1984, \apj, 287, 116

\refindent
Kennicutt, R.~C. 1988, \apj, 334, 144

\refindent
Kennicutt, R.~C. 1992, \apjs, 79, 255

\refindent
Kennicutt, R.~C., Keel, W.~C., \& Blaha, C.~A. 1989, \aj, 97, 1022 (KKB)

\refindent
Kunth, D., \& Sargent, W.~L.~W. 1983, \apj, 273, 81

\refindent
Lehnert, M.~D., \& Heckman, T.~M. 1994, \apj, 426, L27

\refindent
Mathis, J. 1986, \pasp, 98, 995

\refindent
Mathis, J. 1990, \annrev, 28, 37

\refindent
Mazzarella, J.~M., \& Balzano, V.~A. 1986, \apjs, 62, 751

\refindent
McCall, M.~L., Rybski, P.~M., \& Shields, G.~A. 1985, \apjs, 57, 1

\refindent
McGaugh, S.~S. 1991, \apj, 380, 140

\refindent
Miller, B.~W., \& Hodge, P. 1996, \apj, 458, 467

\refindent
Oey, M.~S., \& Kennicutt, R.~C. 1993, \apj, 411, 137

\refindent
Pagel, B.~E.~J., Edmunds, M.~G., Blackwell, D.~E., Chun, M.~S., \& Smith, G. 
1979, \mnras, 189, 95

\refindent
Peimbert, M., \& Torres-Peimbert, S. 1992, \aa, 253, 349

\refindent
Phillips, A.~P., Gondhalekar, P.~M., \& Pettini, M. 1982, \mnras, 200, 687

\refindent
Pogge, R.~W. 1989, \apjs, 71, 433
 
\refindent
Press, W.~H., Flannery, B.~P., Teukolsky, S.~A., \& Vetterling, W.~T. 1986,
Numerical Recipes, the Art of Scientific Computing (Cambridge: Cambridge
Univ. Press)

\refindent
Rand, R.~J. 1996, in The Interplay Between Massive Star Formation, the ISM 
and Galaxy Evolution, ed.  D. Kunth, et al. (Gif-sur-Yvette: Editions 
Fronti\`eres), 515

\refindent
Reynolds, J.~S. 1990, in I.A.U. Symp. 144, The Interstellar Disk/Halo 
Connection in Galaxies, ed. H. Bloemen (Dordrecht: Kluwer), 67
 
\refindent
Roberts, M.~S., \& Haynes, M.~P. 1994, \annrev, 32, 115

\refindent
Ryder, S.~D. 1995, \apj, 444, 610

\refindent
Sandage, A.~R., \& Tammann, G.~A. 1981, A Revised Shapley-Ames Catalog of
Bright Galaxies (Washington, DC: Carnegie Institute of Washington)

\refindent
Shields, G.~A. 1990, \annrev, 28, 525

\refindent
Shields, G.~A., \& Tinsley, B.~M. 1976, \apj, 203, 66

\refindent
Shields, J. C. 1992, \apj, 399, L27

\refindent
Shields, J. C., \& Filippenko, A. V. 1990, \aj, 100, 1034

\refindent
Shields, J.~C, \& Kennicutt, R.~C., Jr. 1995, \apj, in 454, 807

\refindent
Shull, J.~M., \& McKee, C.~F. 1979, \apj, 227, 131

\refindent
Spergel, D.~N., \& Blitz, L. 1992, \nat, 357, 665

\refindent
Stauffer, J.~R. 1981, Ph.D. thesis, University of California, Berkeley
 
\refindent
Stauffer, J.~R. 1982, \apjs, 50, 517

\refindent
Terlevich, R., Melnick, J., Masegosa, J., Moles, M., \& Copetti, M.~V.~F.
1991, \aas, 91, 285

\refindent
Turnrose, B.~E. 1976, \apj, 210, 33

\refindent
van der Hulst, J.~M., Kennicutt, R.~C., Crane, P.~C., \& Rots, A.~H. 1988,
\aa, 195, 38

\refindent
Veilleux, S., \& Osterbrock, D.~E. 1987, \apjs, 63, 295

\refindent
V\'eron, P., \& V\'eron-Cetty, M.-P. 1986, \aa, 161, 145

\refindent
V\'eron, P., V\'eron-Cetty, M.-P., Bergeron, J., \& Zuiderwijk, E.~J. 1981,
\aa, 97, 71

\refindent
V\'\i lchez, J.~M., \& Pagel, B.~E.~J. 1988, \mnras, 231, 257

\refindent
Zaritsky, D., Elston, R., \& Hill, J.~M. 1990, \aj, 99, 1108

\clearpage

\centerline{\bf{Figure Captions}}
\medskip

Fig. 1. ---
Integrated absolute blue magnitudes ($M^0_{B_T}$, corrected for
internal extinction; Ho \etal 1997a) of the host galaxies of \hii\ nuclei 
for ({\it a}) all Hubble types (excluding peculiar classifications), 
({\it b}) types ranging from S0 to Sbc, and ({\it c}) types ranging from Sc 
to I0.  The bins are separated by 0.5 mag.

Fig. 2. ---
H\al\ luminosities [L(H\al)] for ({\it a}) all \hii\ nuclei in our survey 
observed under photometric conditions, ({\it b}) nuclei with types S0--Sbc, 
({\it c}) nuclei with types Sc--I0, ({\it d}) starburst nuclei from Balzano 
(1983), and ({\it e}) first-ranked disk \hii\ regions from Kennicutt (1988).  
Each bin corresponds to 0.25 in logarithmic units.  Our survey luminosities 
were corrected for internal and Galactic reddening, estimated from the 
observed H\al/H\bet\ ratio, using a Galactic extinction curve (see Ho \etal 
1997a).  Since Balzano's individual Balmer decrement values are rather 
uncertain, her luminosities were dereddened using their median H\al/H\bet\ 
ratio, which corresponds to $E(B-V)$ = 0.62 mag.  The L(H\al) values for the 
disk \hii\ regions represent the average luminosity of the three first-ranked 
regions as tabulated by Kennicutt, who only corrected for Galactic reddening.  
We dereddened the luminosities by $E(B-V)$ = 0.29 mag, the median internal 
reddening of our compilation of disk \hii\ regions.

Fig. 3. ---
H\bet\ emission equivalent widths for ({\it a}) all \hii\ nuclei in our 
survey, ({\it b}) nuclei with types S0--Sbc, ({\it c}) nuclei with types 
Sc--I0, and ({\it d}) the sample of disk \hii\ regions of McCall \etal 
(1985).  The bins are separated by 5 \AA, and the last bin contains all 
objects with EW(H\bet) $>$ 99 \AA.

Fig. 4. ---
Distribution of internal reddening [$E(B-V)_{\rm int}$] for
({\it a}) all \hii\ nuclei in our survey, ({\it b}) nuclei with types S0--Sbc,
({\it c}) nuclei with types Sc--I0, and ({\it d}) the sample of disk
\hii\ regions of McCall \etal (1985).  The bins are separated by 0.1 mag.  
We estimated the internal reddening in our measurements from the observed 
H\al/H\bet\ ratio (see Ho \etal 1997a).

Fig. 5. ---
Electron densities ($n_e$) derived from the \sii\ \lamb 6716/\lamb 6731 ratio 
for ({\it a}) all \hii\ nuclei in our survey, ({\it b}) nuclei with types 
S0--Sbc, ({\it c}) nuclei with types Sc--I0, and ({\it d}) the sample of disk 
\hii\ regions of McCall \etal (1985).  The bins are separated by 50 \cc.

Fig. 6. ---
Extinction-corrected \oiii\ \lamb 5007/H\bet\ ratios for ({\it a}) all \hii\ 
nuclei in our survey, ({\it b}) nuclei with types S0--Sbc, ({\it c}) nuclei 
with types Sc--I0, ({\it d}) starburst nuclei from Balzano (1983), and 
({\it e}) disk \hii\ regions from several sources.  The bins are separated 
by units of 0.2.

Fig. 7. ---
Diagnostic diagram, adapted from Veilleux \& Osterbrock (1987),
plotting log \oiii\ \lamb 5007/H\bet\ versus log \nii\ \lamb 6583/H\al.
The symbols have the following meaning: {\it solid circles} = disk \hii\
regions, {\it open circles} = \hii\ nuclei, and {\it asterisks} = starburst
nuclei.  The {\it dashed curve} denotes the theoretical models of disk
\hii\ regions by McCall \etal (1985).  The calculations assume that the
effective temperature of the ionizing stars decreases with increasing
oxygen abundance, ranging from $T_{eff}$ = 47,000 to 38,500 K as
(O/H)/(O/H)$_{\odot}$ varies from 0.5 to 3.5 ({\it top} to {\it bottom}).
Also shown are the photoionization models of Shields \& Kennicutt (1995) which
incorporate the effects of dust ({\it dotted curves}); the top curve was 
calculated for $T_{eff}$ = 45,000 K and the bottom one for $T_{eff}$ = 
38,000 K.  The models assume a constant density and ionization parameter, 
with abundances varying between 0.1 and 5 times the solar values from the 
upper left to the bottom right of each curve.
The unusual object IC 2574 (see \S\ 4.2) is labeled.

Fig. 8. ---
Extinction-corrected \nii\ \lamb 6583/H\al\ ratio for ({\it a}) all \hii\ 
nuclei in our survey, ({\it b}) nuclei with types S0--Sbc, ({\it c}) nuclei 
with types Sc--I0, ({\it d}) starburst nuclei from Balzano (1983), and 
({\it e}) disk \hii\ regions from several sources.  The bins are separated
by units of 0.05.

Fig. 9. ---
Diagnostic diagram plotting log \oiii\ \lamb 5007/H\bet\
versus log \nii\ \lamb 6583/H\al\ for ({\it solid circles}) early-type
(S0--Sbc) \hii\ nuclei and ({\it open circles}) late-type (Sc--I0) \hii\
nuclei. The curve denotes the theoretical models of disk \hii\ regions by
McCall \etal (1985) described in the caption of Figure 7.  
The unusual object IC 2574 (see \S\ 4.2) is labeled.

Fig. 10. ---
Diagnostic diagram plotting log \oiii\ \lamb 5007/H\bet\
versus log \sii\ \lamb\lamb 6716, 6731/H\al.  The symbols have the same
meaning as in Figure 7.

Fig. 11. ---
Extinction-corrected \sii\ \lamb\lamb 6716, 6731/H\al\ ratio for ({\it a}) all 
\hii\ nuclei in our survey, ({\it b}) nuclei with types S0--Sbc, ({\it c}) 
nuclei with types Sc--I0, ({\it d}) starburst nuclei from Balzano (1983), 
and ({\it e}) disk \hii\ regions from several sources.  The bins are separated
by units of 0.05.

Fig. 12. ---
Diagnostic diagram plotting log \oiii\ \lamb 5007/H\bet\ 
versus log \sii\ \lamb\lamb 6716, 6731/H\al\ for ({\it solid circles}) 
early-type (S0--Sbc) \hii\ nuclei and ({\it open circles}) late-type (Sc--I0) 
\hii\ nuclei. The curve denotes the theoretical models of disk \hii\ regions by 
McCall \etal (1985) described in the caption of Figure 7.

Fig. 13. ---
Diagnostic diagram plotting log \oiii\ \lamb 5007/H\bet\ 
versus log \oi\ \lamb 6300/H\al.  The symbols have the following meaning: 
{\it open triangles} = disk \hii\ regions, {\it solid circles} = early-type 
(S0--Sbc) \hii\ nuclei, {\it open circles} = late-type (Sc--I0) \hii\ nuclei, 
and {\it asterisks} = starburst nuclei.  The curves denote the theoretical 
models of disk \hii\ regions described in the caption of Figure 7.

Fig. 14. ---
Extinction-corrected \oi\ \lamb 6300/H\al\ ratio for ({\it a}) all \hii\ 
nuclei in our survey, ({\it b}) nuclei with types S0--Sbc, ({\it c}) nuclei 
with types Sc--I0, ({\it d}) starburst nuclei from Balzano (1983), and 
({\it e}) disk \hii\ regions from several sources.  The bins are separated
by units of 0.01.

Fig. 15. ---
The line ratios \sii\ \lamb\lamb 6716, 6731/H\al\ and \oi\ \lamb 6300/H\al\ 
correlate well with each other.  {\it Solid circles} denote early-type 
(S0--Sbc) \hii\ nuclei, and {\it open circles} denote late-type (Sc--I0) 
\hii\ nuclei.

\clearpage
\begin{figure}
\figurenum{1}
\plotone{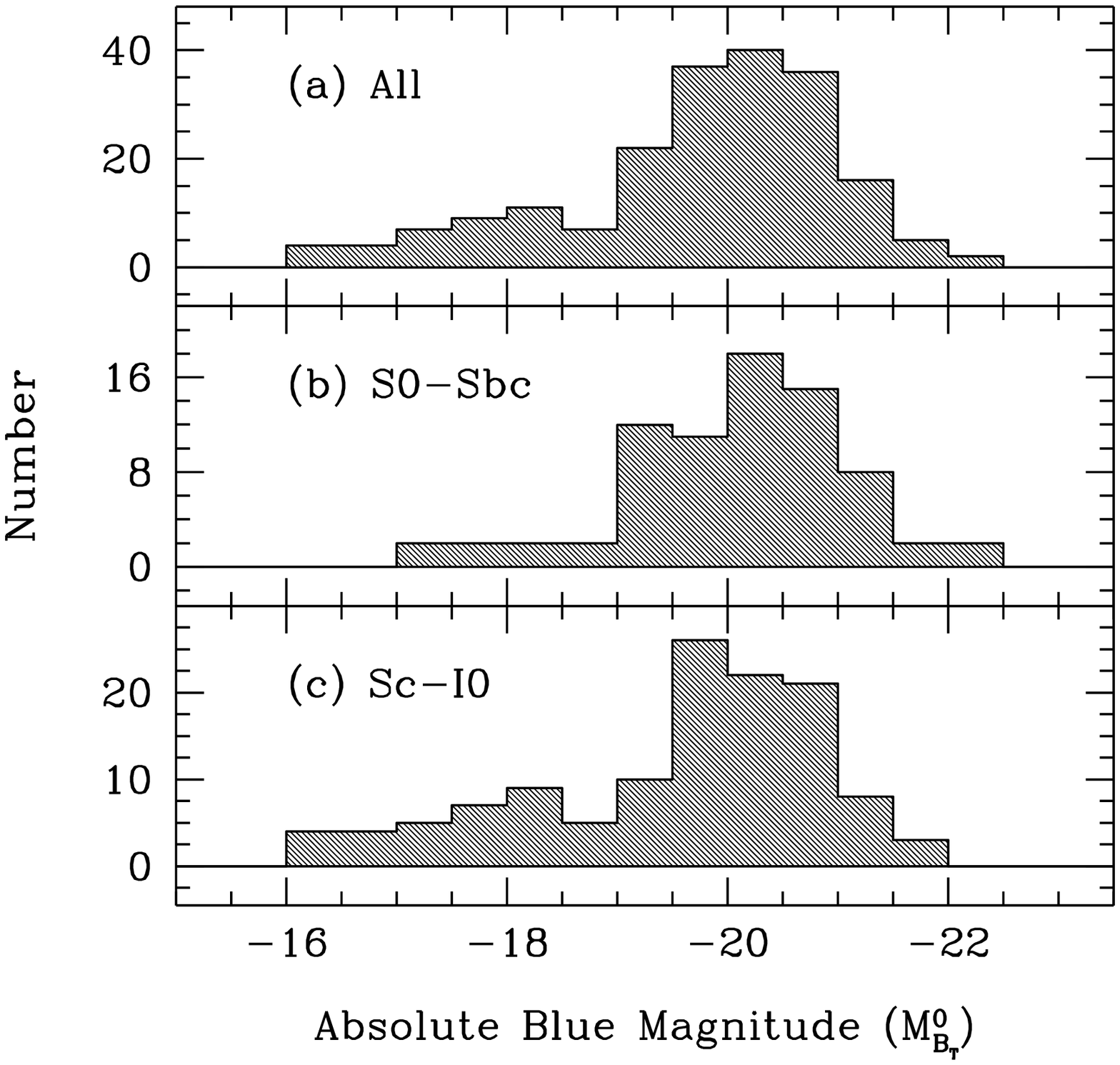}
\caption{}
\end{figure}

\clearpage
\begin{figure}
\figurenum{2}
\plotone{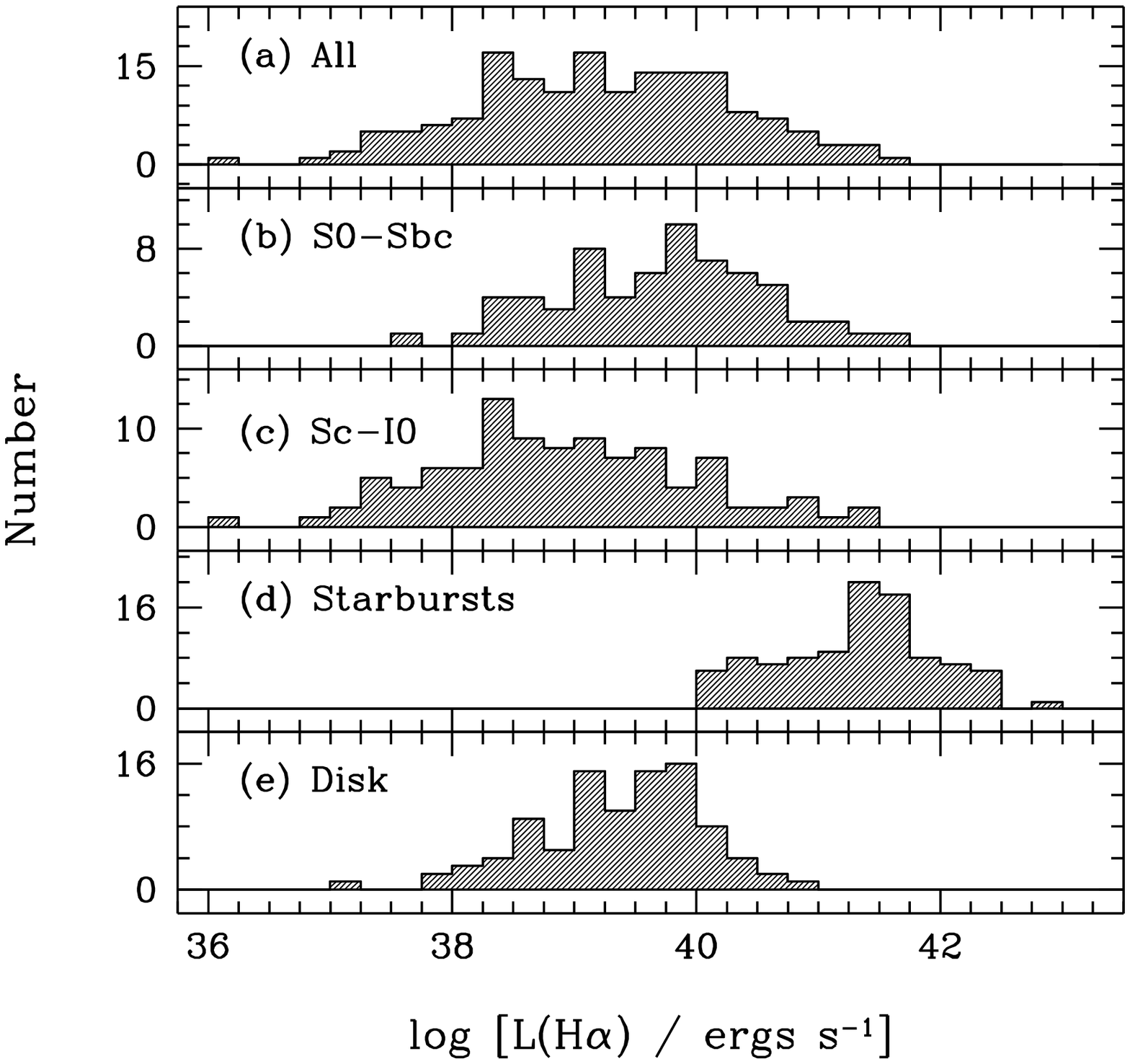}
\caption{}
\end{figure}

\clearpage
\begin{figure}
\figurenum{3}
\plotone{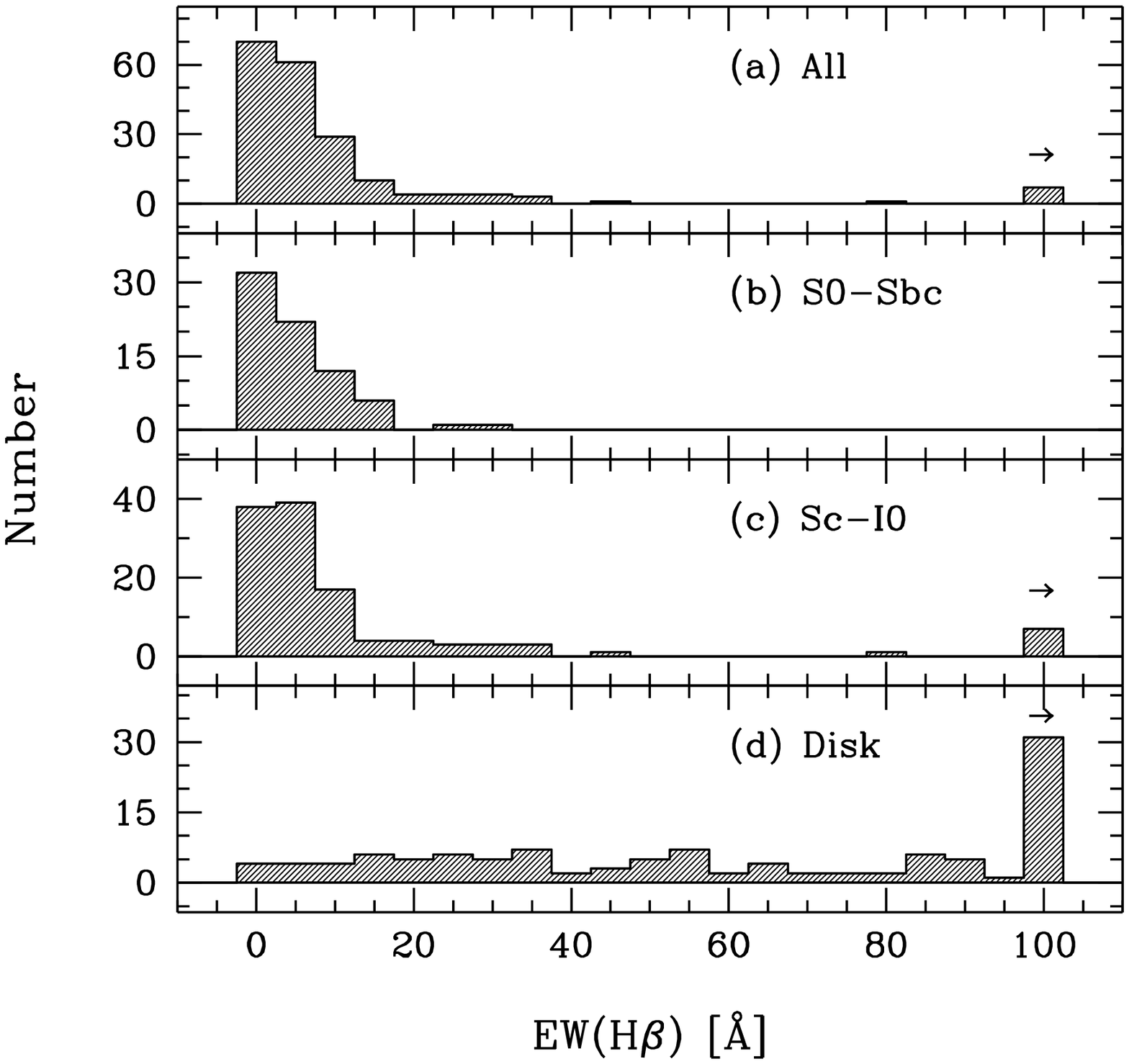}
\caption{}
\end{figure}

\clearpage
\begin{figure}
\figurenum{4}
\plotone{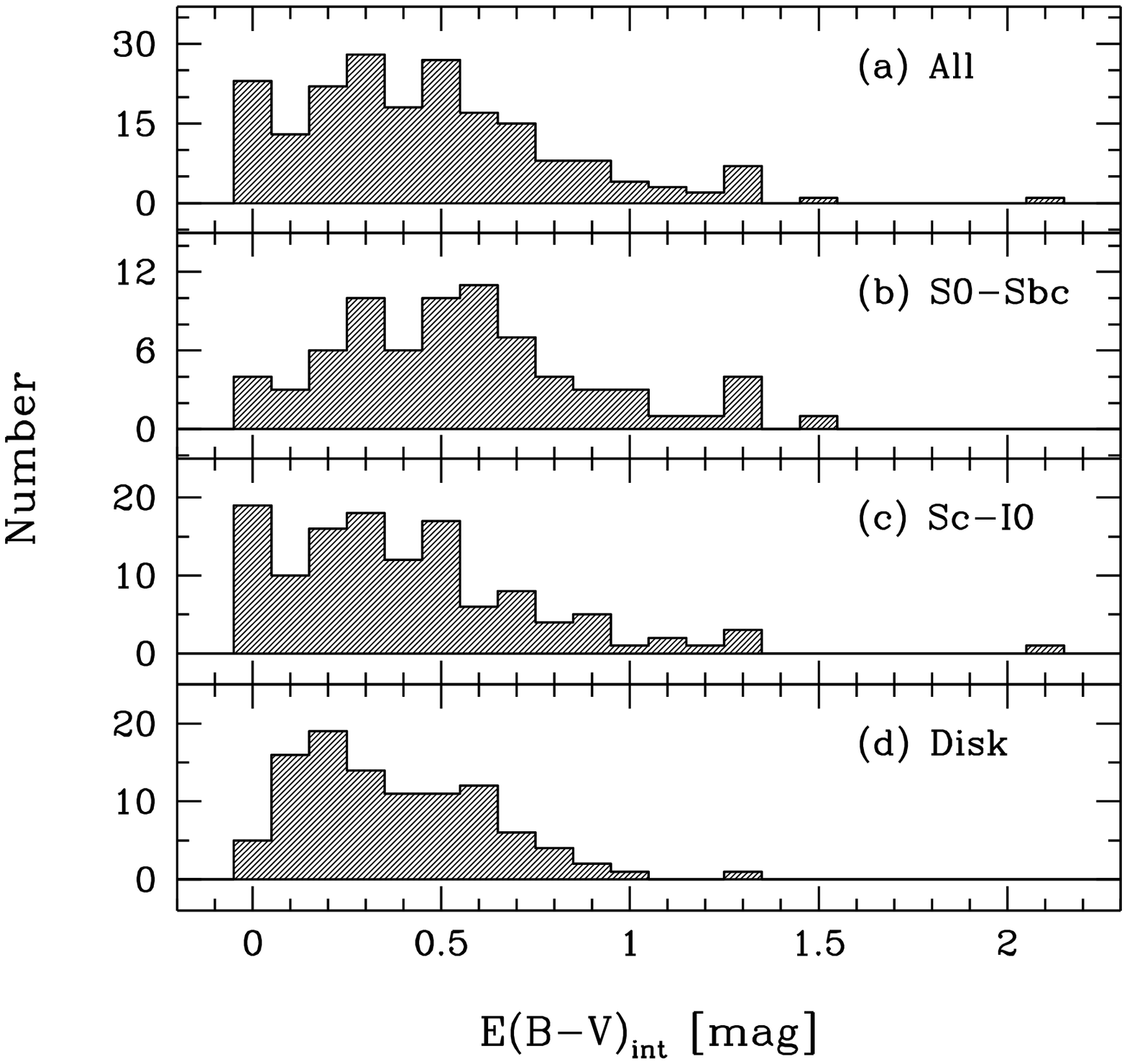}
\caption{}
\end{figure}

\clearpage
\begin{figure}
\figurenum{5}
\plotone{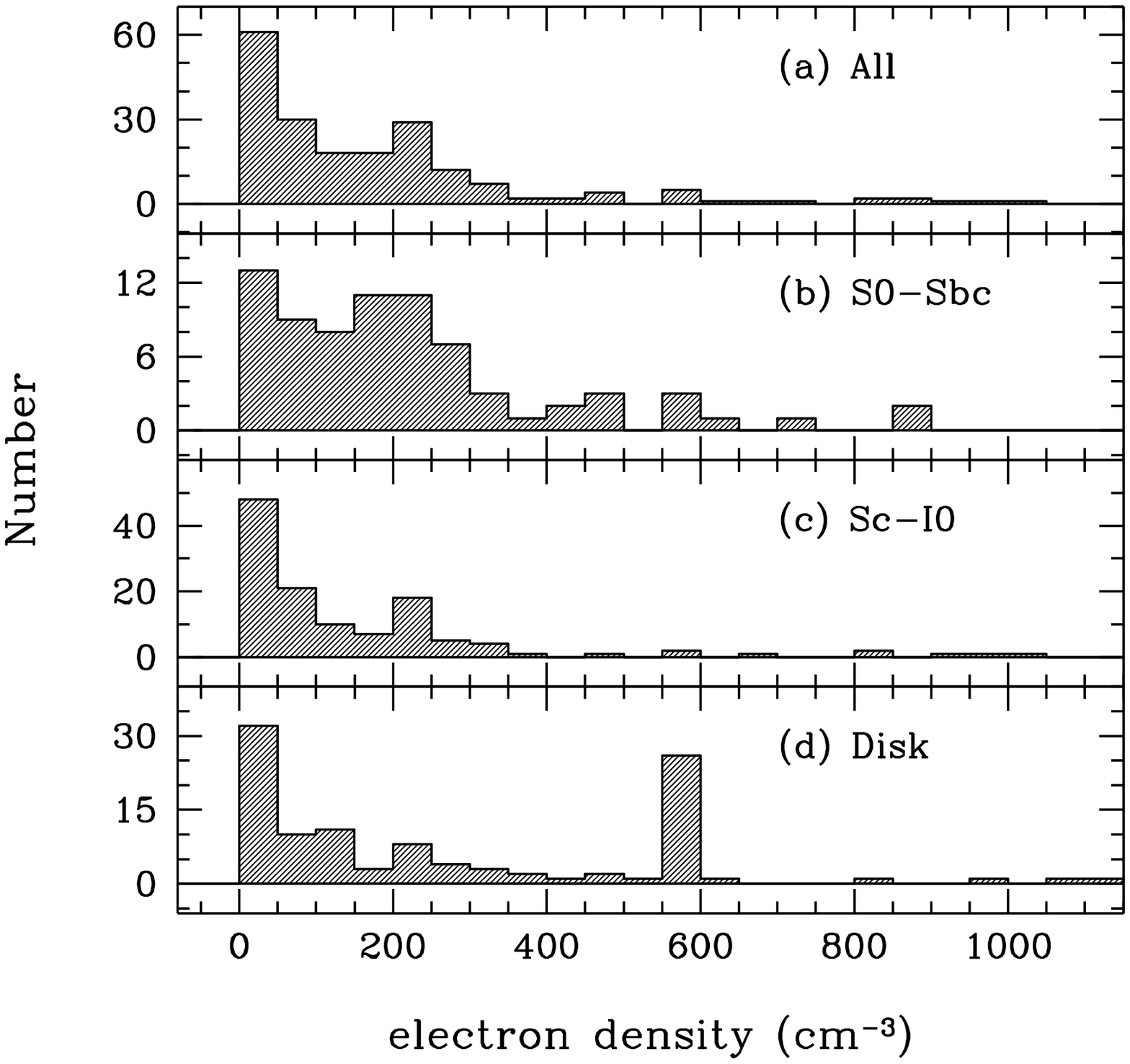}
\caption{}
\end{figure}

\clearpage
\begin{figure}
\figurenum{6}
\plotone{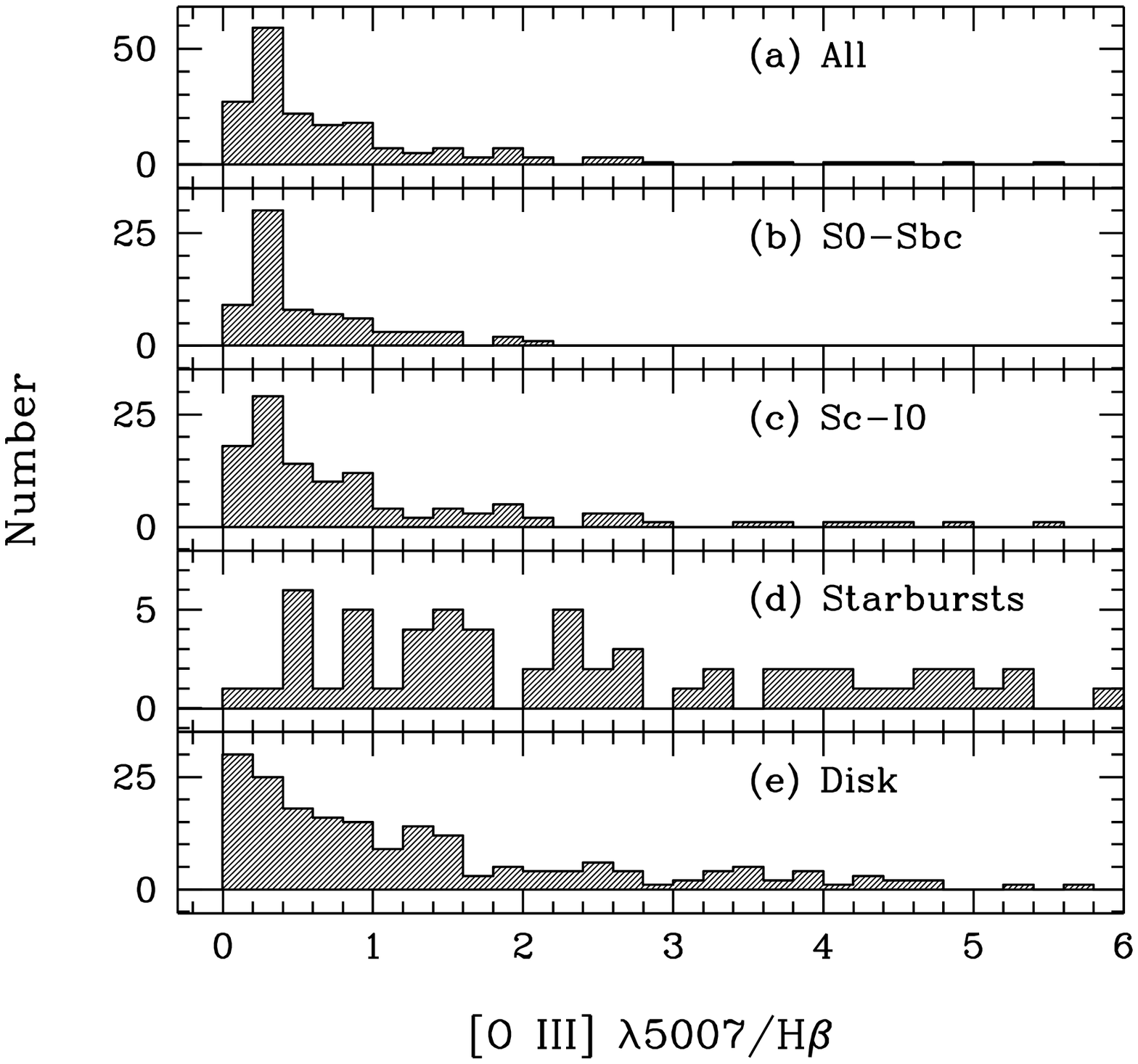}
\caption{}
\end{figure}

\clearpage
\begin{figure}
\figurenum{7}
\plotone{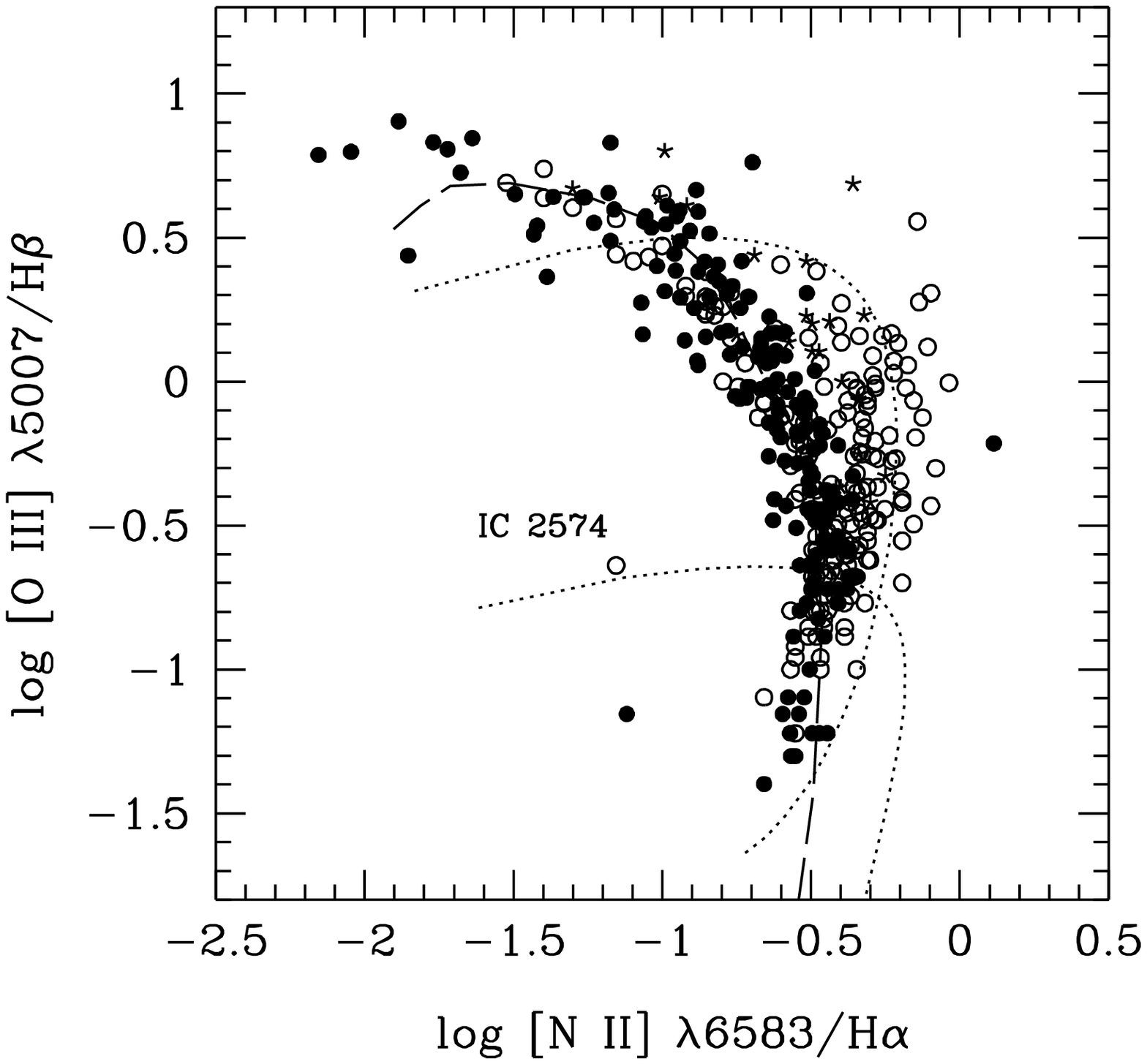}
\caption{}
\end{figure}

\clearpage
\begin{figure}
\figurenum{8}
\plotone{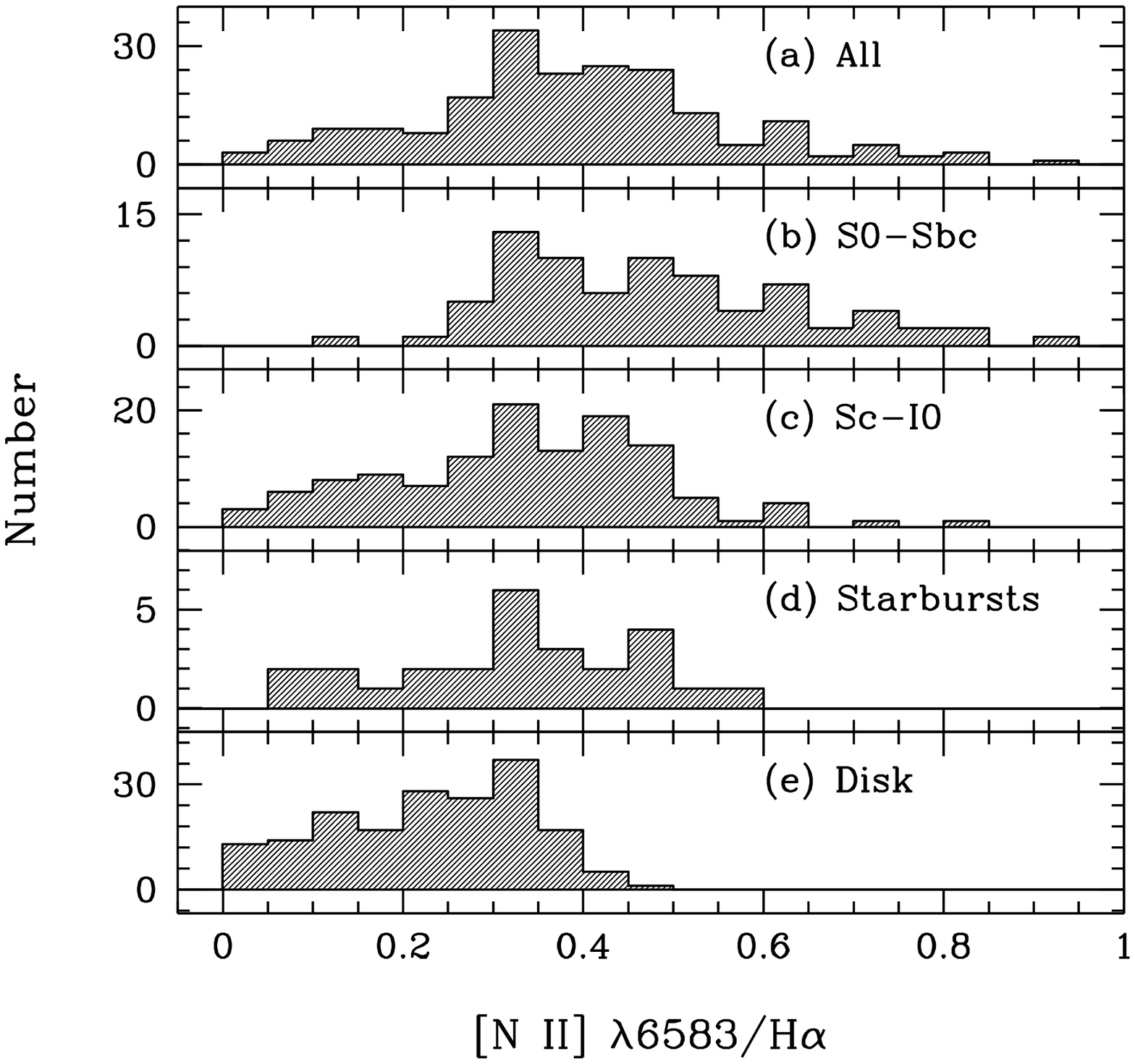}
\caption{}
\end{figure}

\clearpage
\begin{figure}
\figurenum{9}
\plotone{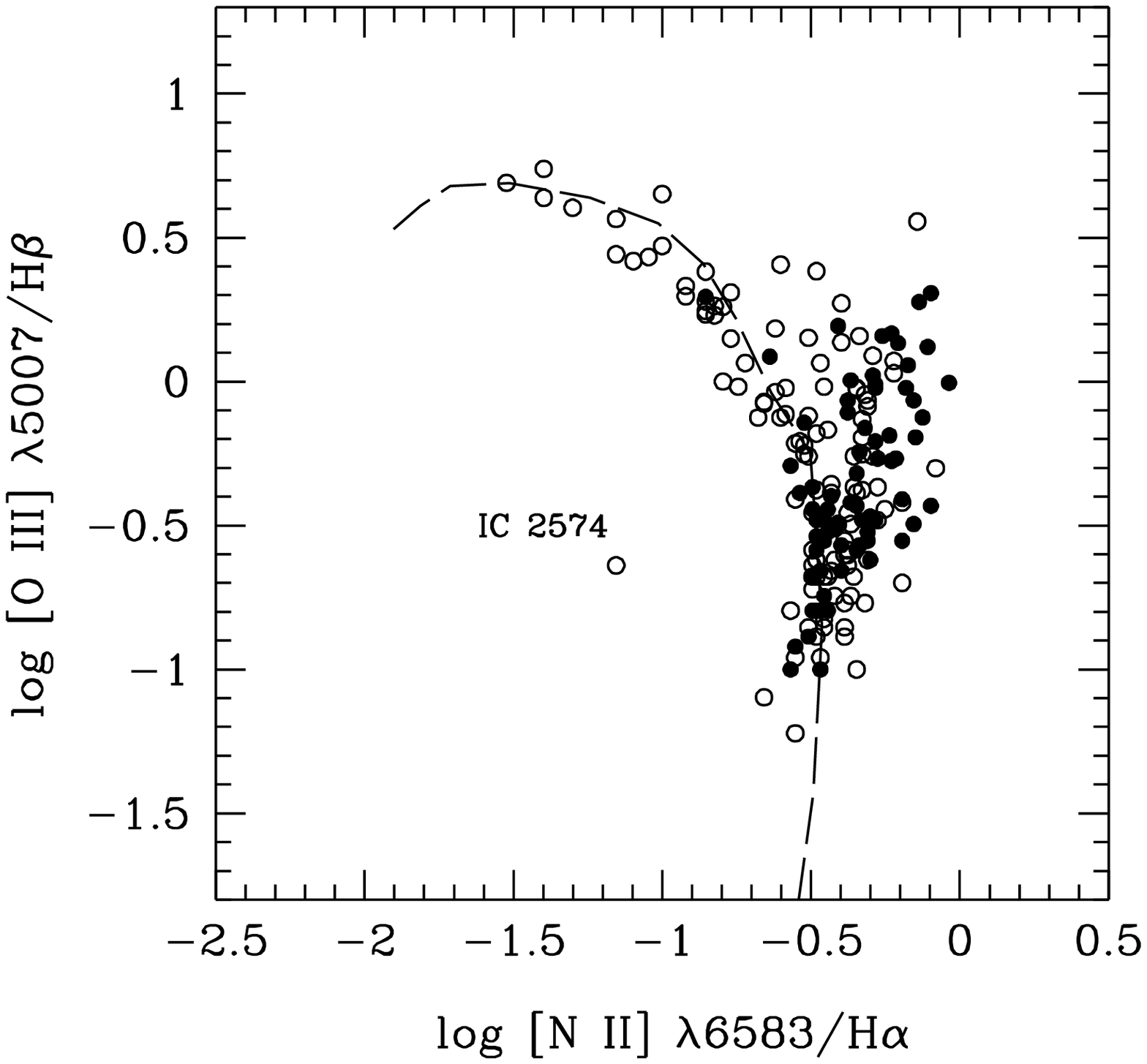}
\caption{}
\end{figure}

\clearpage
\begin{figure}
\figurenum{10}
\plotone{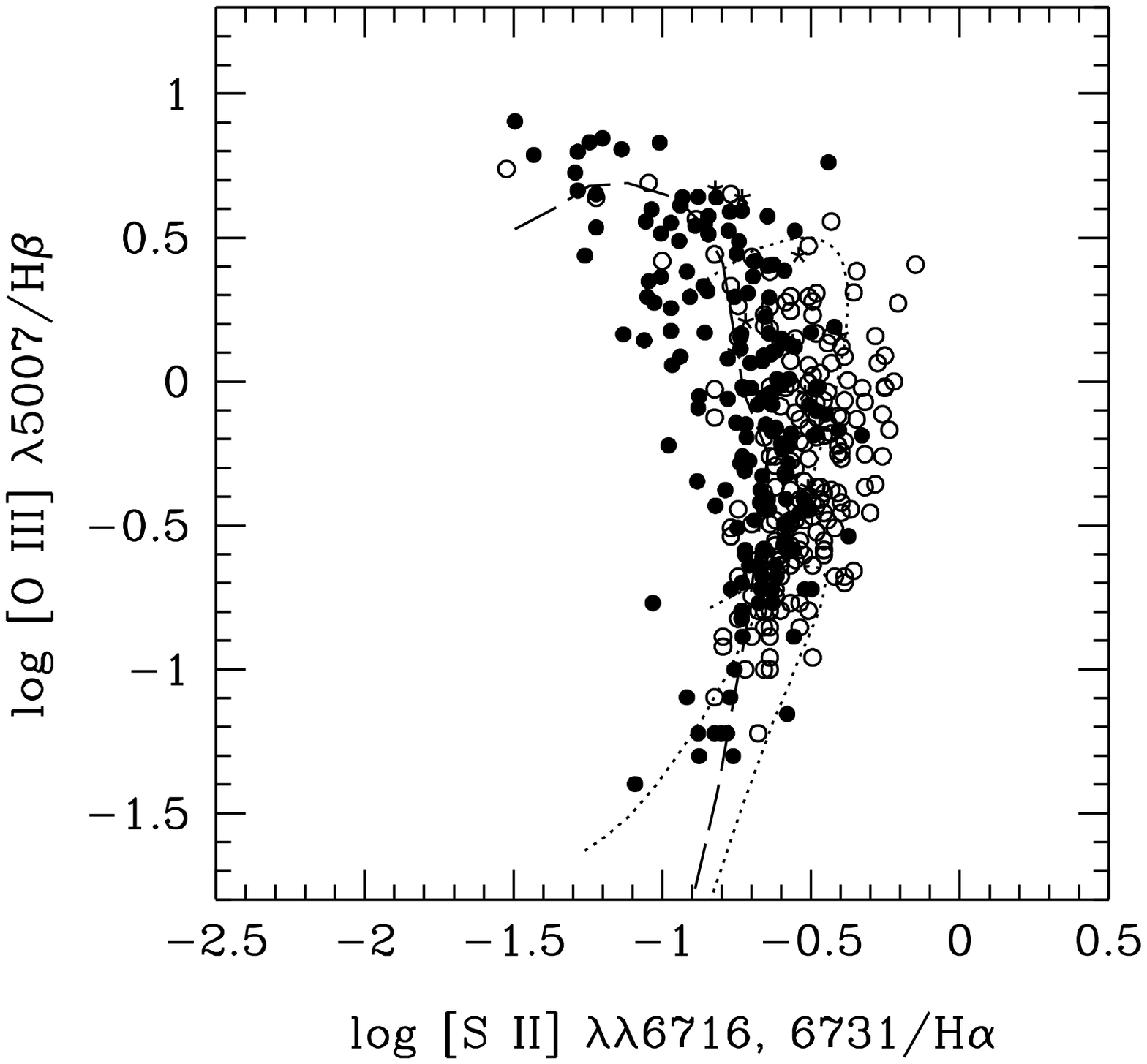}
\caption{}
\end{figure}

\clearpage
\begin{figure}
\figurenum{11}
\plotone{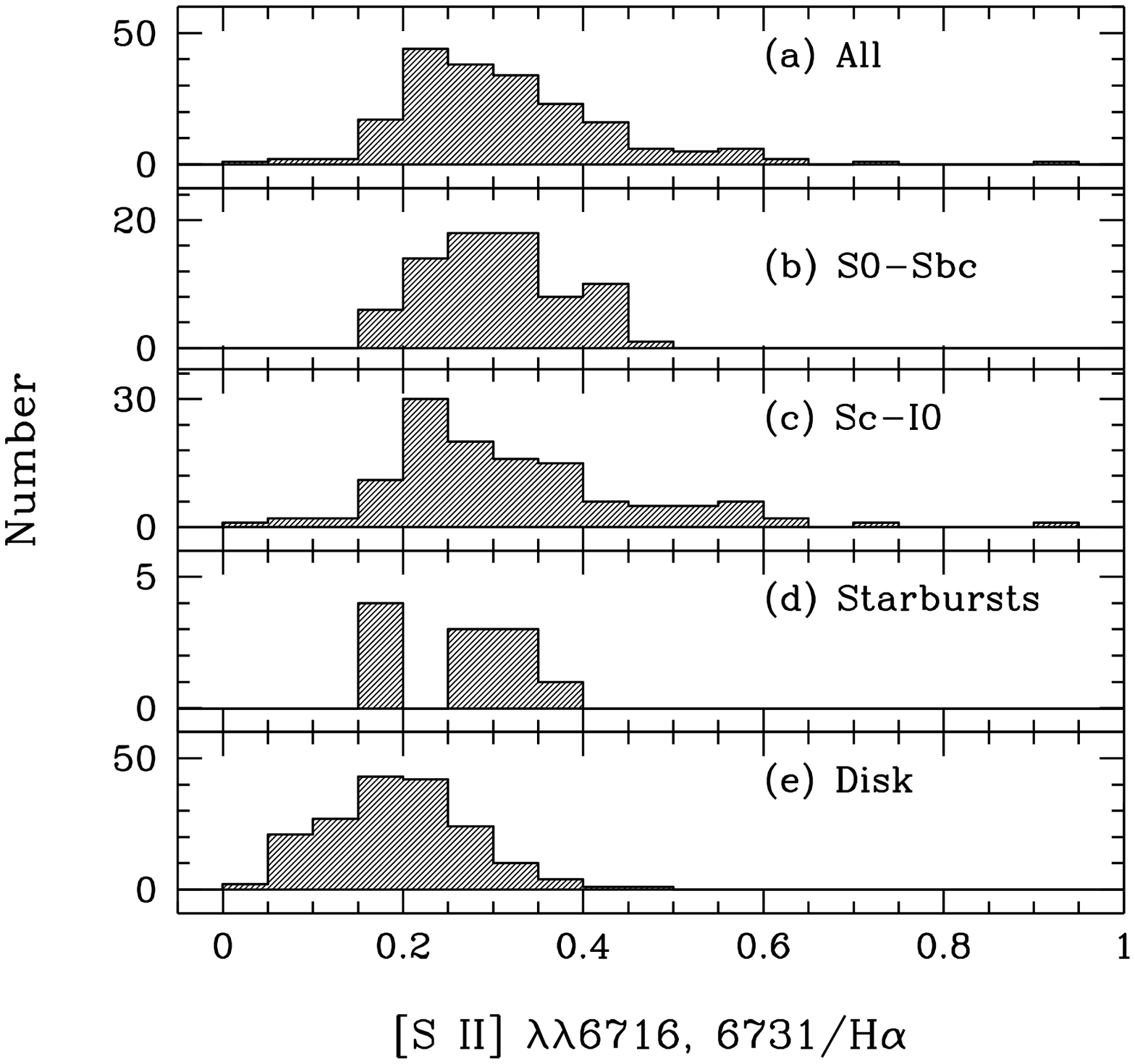}
\caption{}
\end{figure}

\clearpage
\begin{figure}
\figurenum{12}
\plotone{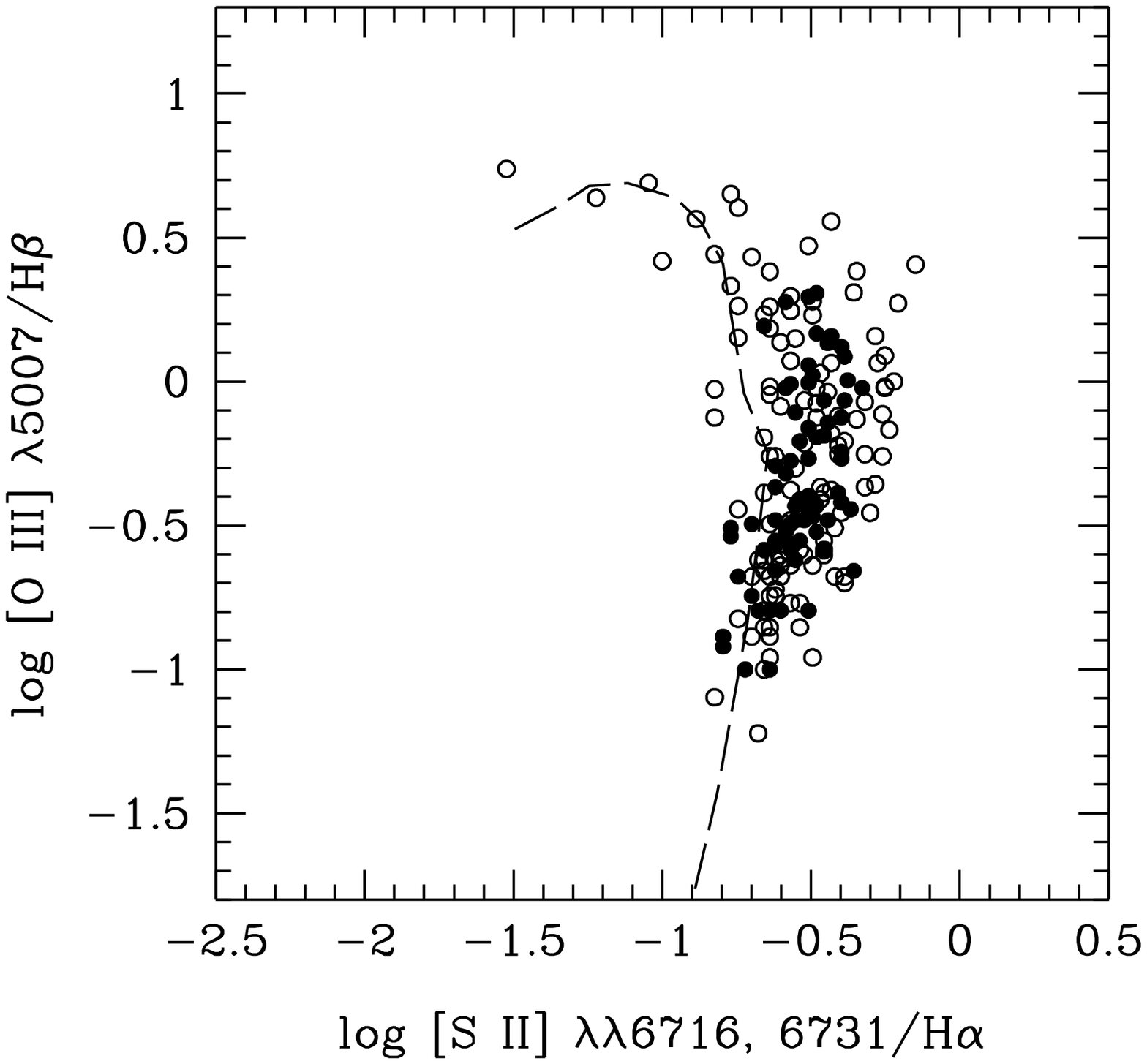}
\caption{}
\end{figure}

\clearpage
\begin{figure}
\figurenum{13}
\plotone{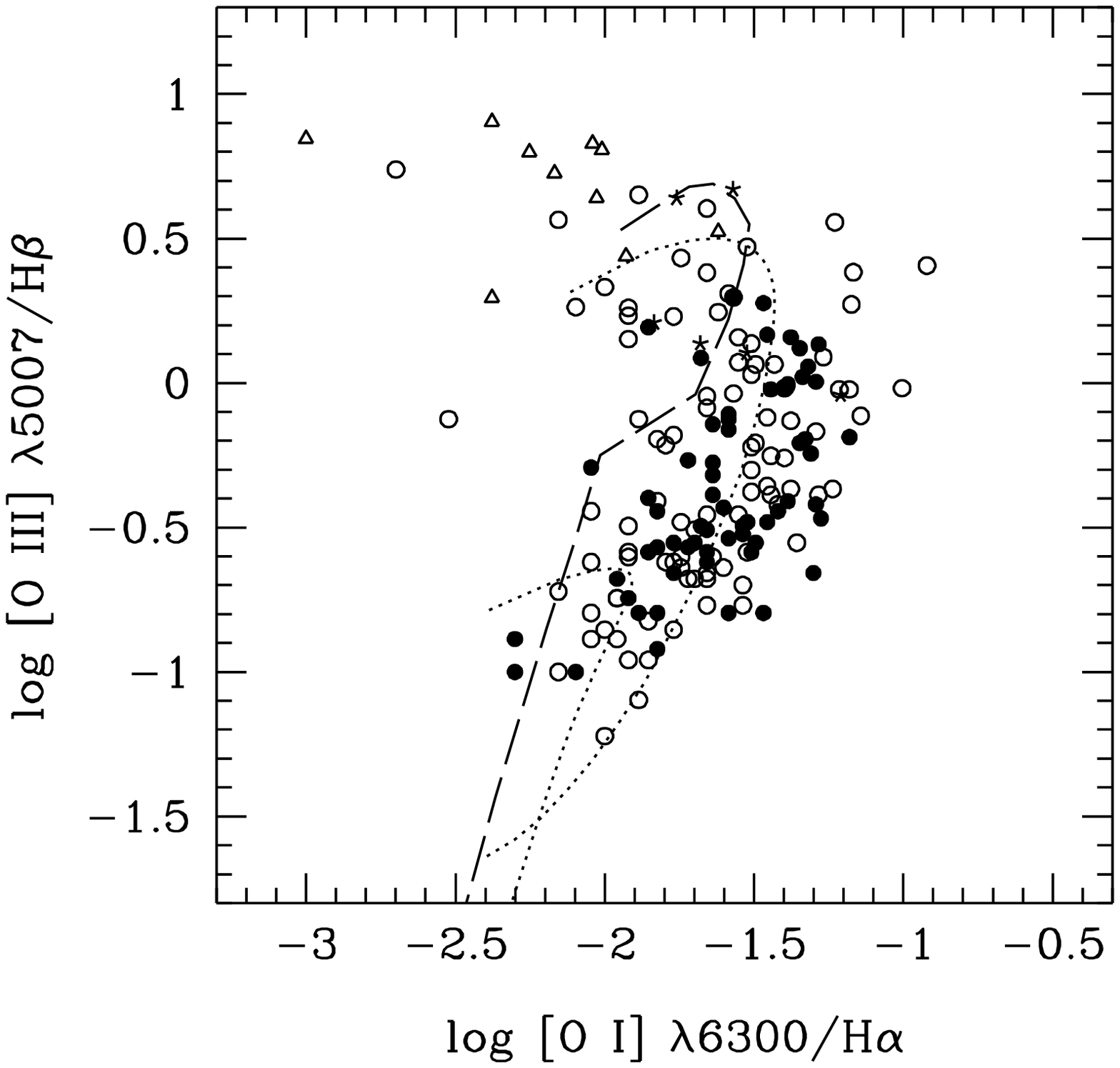}
\caption{}
\end{figure}

\clearpage
\begin{figure}
\figurenum{14}
\plotone{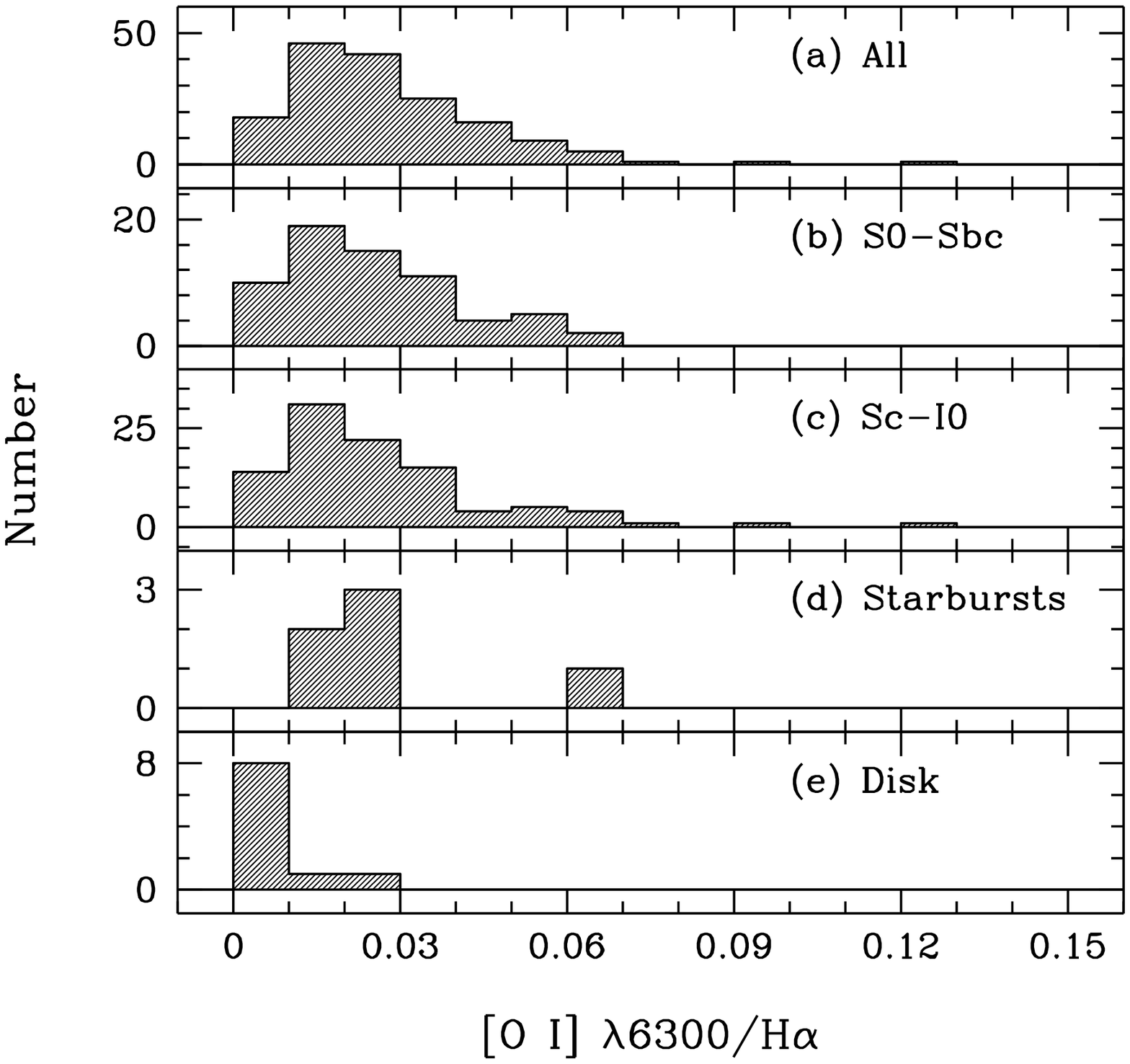}
\caption{}
\end{figure}

\clearpage
\begin{figure}
\figurenum{15}
\plotone{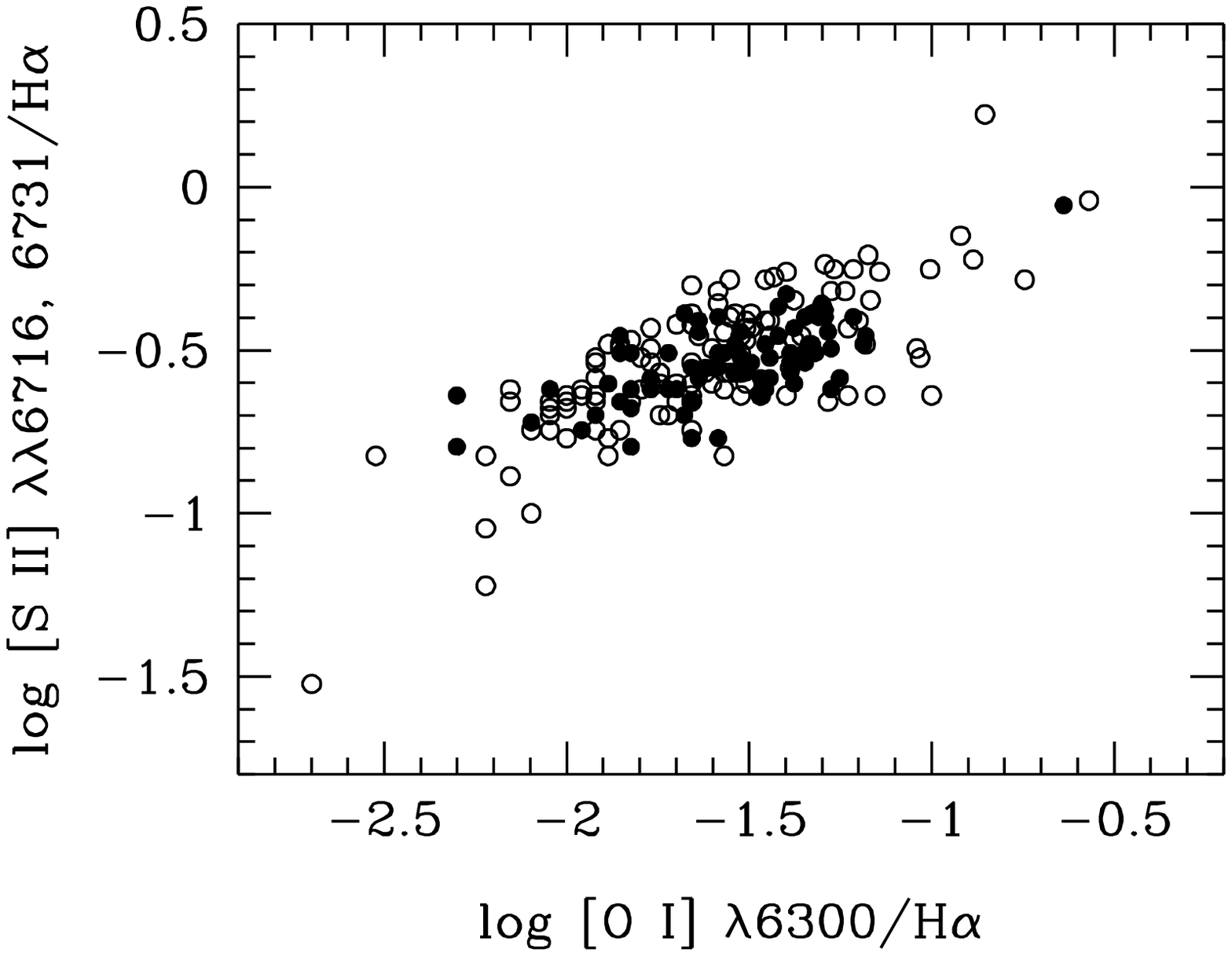}
\caption{}
\end{figure}

\end{document}